\documentclass[aps,prd,preprint,showpacs,nofootinbib,floatfix]{revtex4}

\usepackage{bm}
\usepackage{graphicx}%
\usepackage{epsf}
\usepackage{epsfig}

\usepackage{ulem}

 \usepackage{amsmath}
\usepackage{amsfonts,amssymb}

\newcommand{\be}{\begin{equation}}
\newcommand{\ee}{\end{equation}}
\newcommand{\bea}{\begin{eqnarray}}
\newcommand{\eea}{\end{eqnarray}}

\newcommand{\ggcf}{\frac{g^2 C_F}{16 \, \pi^2}\; }
\def\smn{{\sigma_{\mu\nu}}}

\usepackage{dsfont}
\usepackage{slashed}
\usepackage{booktabs}
\usepackage{placeins}
\usepackage{wasysym}

\begin{document}

\vspace*{1.75cm}

\title{Renormalization of local quark-bilinear operators \\ for
   $\mathbf{N_f=3}$ flavors of SLiNC fermions}

\author{M. Constantinou$^{1}$, R. Horsley$^{2}$,
  H. Panagopoulos$^{1}$, H. Perlt$^{3}$,
  P.E.L. Rakow$^{4}$, G. Schierholz$^{5}$, A. Schiller$^{3}$ and
  J.M. Zanotti$^{6}$} 

\affiliation{\vspace*{0.5cm}
$^1$ Department of Physics, University of Cyprus, Nicosia CY-1678,
Cyprus\\
$^2$ School of Physics and Astronomy, University of Edinburgh, Edinburgh
EH9 3JZ, United Kingdom\\  
$^3$ Institut f\"ur Theoretische Physik, Universit\"at Leipzig, 04103
Leipzig, Germany \\  
$^4$ Theoretical Physics Division, Department of Mathematical Sciences,
University of Liverpool, Liverpool L69 3BX, United Kingdom\\  
$^5$ Deutsches Elektronen-Synchrotron DESY, 22603 Hamburg, Germany \\ 
$^6$ CSSM, School of Chemistry and Physics, University of Adelaide,
Adelaide SA 5005, Australia\\
}

\renewcommand{\baselinestretch}{1.5}\normalsize

\begin{abstract}
The renormalization factors of local quark-bilinear operators are
computed non-perturbatively for $N_f=3$ flavors of SLiNC fermions, with
emphasis on the various procedures for the chiral and continuum
extrapolations. The simulations are performed at a lattice
spacing $a=0.074$ fm, and for five values of the pion mass in the range
of 290-465 MeV, allowing a safe and stable chiral extrapolation. 
Emphasis is given in the subtraction of the well-known pion pole
which affects the renormalization factor of the pseudoscalar current.
We also compute the inverse propagator and the Green's functions of
the local bilinears to one loop in perturbation theory. We investigate
lattice artifacts by computing them perturbatively to second order as
well as to all orders in the lattice spacing. The renormalization
conditions are defined in the RI$'$-MOM scheme, for both the perturbative and
non-perturbative results. The renormalization factors, obtained at
different values of the renormalization scale, are translated to
the ${\overline{\rm MS}}$ scheme and are evolved perturbatively to 2 GeV.
Any residual dependence on the initial renormalization scale is
eliminated by an extrapolation to the continuum limit. We also study the
various sources of systematic errors.

Particular care is taken in correcting the non-perturbative estimates
by subtracting lattice artifacts computed to one loop perturbation
theory using the same action. We test two different methods, by
subtracting either the ${\cal O}(g^2\,a^2)$ contributions, or the
complete (all orders in $a$) one-loop lattice artifacts.

\end{abstract}


\preprint{\vtop{\hbox{ADP-14-16/T874} 
    \hbox{DESY 14-145}
    \hbox{Edinburgh 2014/16}\hbox{LTH 1018}}}

\maketitle

\section{Introduction}

To make contact between lattice QCD simulation results and
phenomenological numbers, which usually are given   
in the $\overline{\mathrm {MS}}$ scheme, we need to compute
renormalization factors relating the bare lattice operators to
renormalized ones.
This requires a non-perturbative method, because low-order lattice
perturbation theory is unreliable at present couplings. One such
method is the `Regularization Independent Momentum' (RI-MOM)
renormalization scheme~\cite{Martinelli:1994ty}, which mimics the
renormalization 
procedure used in continuum perturbation theory. In practical
applications a variant of the RI-MOM scheme is preferred, the
so-called RI$'$-MOM scheme, which differs from the RI-MOM scheme
only in the quark field renormalization factor. The results can be
converted to the $\overline{\mathrm {MS}}$ scheme using continuum
perturbation theory.

The method involves comparing lattice calculations of off-shell
Green's functions with continuum perturbation theory results. This 
matching will work best at large virtualities or short distances,
where the running coupling constant is small, and the effects of
non-perturbative phenomena, such as chiral symmetry breaking, have
died away. A drawback of the method is that discretization effects
grow rapidly at short distances. It is thus desirable to remove the
discretization errors in the off-shell lattice Green's functions
before making the comparison with the continuum. 

Two approaches have been pursued. In~\cite{Gockeler:2010yr} the full
one-loop perturbative correction has been subtracted from the Green's
functions for operators with up to one covariant derivative, for the
plaquette action and $N_f=2$ flavors of clover fermions. This
procedure becomes very elaborate for higher dimensional operators and
more complex actions. An alternative approach is to restrict oneself
to one loop corrections of $O(a^2)$. The relevant expressions have
been computed in~\cite{Constantinou:2009tr} for a variety of operators
and actions. In~\cite{Constantinou:2013ada} we have compared the
results of both methods using the Wilson action; we found agreement
for that action to be better than $1\%$ for local operators and better
than $2\%$ for operators with one covariant derivative.  

In this work we compute renormalization factors for $N_f=3$
flavors of SLiNC (Stout Link Non-perturbative Clover)
fermions~\cite{Cundy:2009yy,Bietenholz:2011qq} and
local quark-bilinear operators, studying the effects of one-loop
$O(a^2)$ subtraction, as well as the complete subtraction to one
loop. The particular clover action used here has a single iterated
mild stout smearing for the hopping terms, together with thin links
for the clover term. For the gauge fields we are using the tree-level
Symanzik action.

The paper is organized as follows: In Sec.~\ref{sec2} we explain our 
method of non-perturbative renormalization and its numerical
implementation, along with the data sets used in this work and a
precise determination of the lattice scale. In Sec.~\ref{sec3} we
briefly outline our perturbative computation and the different methods
of subtraction. Then, in Sec.~\ref{sec4} we present our
non-perturbative results in the RI$'$-MOM scheme, the chiral
extrapolation and the conversion to the $\overline{\rm MS}$-scheme.
The continuum extrapolation appears in Sec.~\ref{sec5}, where in
Subsec.~\ref{sec5a} we give our final non-perturbative estimates in
the schemes $\overline{\rm MS}$ at 2GeV, Renormalization Group
Invariant (RGI), and RI$'$-MOM at a scale of 1/a.
The quality of the data upon subtraction of the perturbative terms is
demonstrated in Subsec.~\ref{sec5b}, and finally, in Sec.~\ref{sec6} we
present our final results and
conclude. We also include three Appendices: A. A summary of our
perturbative results, B. The $\beta$-function and anomalous
dimensions that are necessary for the conversion to the 
$\overline{\rm MS}$ and RGI schemes, and C. An alternative fitting for the pion pole subtraction.

\vspace{-0.3cm}
\section{Method and numerical implementation}
\label{sec2}
\vspace{-0.3cm}
\begin{table}[!h]
\begin{center}
\begin{tabular}{|c|c|c|c|}\hline
Operator & Lorentz Structure & Representation & Operator Basis\\ \hline
$\mathcal{O}^S$ & $\bar{q} q$ & $\tau_1^{(1)}$ &  $\mathcal{O}^S$ \\ 
$\mathcal{O}^P$ & $\bar{q} \gamma_5 q$ & $\tau_1^{(1)}$ & $\mathcal{O}^P$  \\ 
$\mathcal{O}_\mu^V$ & $\bar{q} \gamma_\mu q$ & $\tau_1^{(4)}$ &
$\mathcal{O}_1^V$, $\mathcal{O}_2^V$, $\mathcal{O}_3^V$, $\mathcal{O}_4^V$ \\  
$\mathcal{O}_\mu^A$ & $\bar{q} \gamma_\mu\gamma_5 q$ & $\tau_1^{(4)}$
&  $\mathcal{O}_1^A$, $\mathcal{O}_2^A$, $\mathcal{O}_3^A$,
$\mathcal{O}_4^A$ \\
$\mathcal{O}_{\mu\nu}^T$ & $\bar{q} \sigma_{\mu\nu} q$ &
$\tau_1^{(6)}$ & $\mathcal{O}_{12}^T$, $\mathcal{O}_{13}^T$,
$\mathcal{O}_{14}^T$, $\mathcal{O}_{23}^T$, $\mathcal{O}_{24}^T$,
$\mathcal{O}_{34}^T$ \\ \hline
\end{tabular}
\end{center}
\vskip -0.4cm
\caption{The operators under study and their representations under the hypercubic
  group H(4)~\cite{Gockeler:1996mu}.}   
\label{tab1}
\end{table}

\vskip -0.2cm
The operators which we study in this
paper are listed in Table~\ref{tab1}, along with their representations
under the hypercubic group $H(4)$~\cite{Gockeler:1996mu}. We work on lattices with
spacing $a$ and volume $V$, gauge fixed to Landau gauge. Starting from
the Green's function
\begin{equation} \label{Gdef}
 G_{\alpha\beta} (p) = \frac{a^{12}}{V} \sum_{x,y,z} {\rm e}^{- {\mathrm i} 
      p\cdot (x-y) } \langle q_\alpha (x) \mathcal{O} (z) \bar{q}_\beta (y)
 \rangle \,,
\end{equation}
$q=u, d$ or $s$, with operator insertion $\mathcal{O}$, we obtain the
vertex  function (or amputated Green's function) 
\begin{equation}
\Gamma (p) = S^{-1} (p) G(p) S^{-1} (p) \,, 
\end{equation}
where  
\begin{equation} \label{Sdef}
 S_{\alpha\beta} (p) = \frac{a^8}{V} \sum_{x,y} {\rm e}^{- {\mathrm i} 
      p \cdot (x-y) } \langle q_\alpha (x) \bar{q}_\beta (y) \rangle 
\end{equation}
denotes the quark propagator. 
Following~\cite{Martinelli:1994ty,Gockeler:2010yr,Constantinou:2013ada}, 
we define the renormalized vertex function by  
\begin{equation}
\Gamma_{\mathrm R} (p) = Z_q^{-1} Z\, \Gamma (p)
\end{equation}
and fix the renormalization factor $Z$ by imposing the
renormalization condition  
\begin{equation} \label{defz}
 \mbox{\small $\frac{1}{12}$} {\rm Tr} \left[ \Gamma_{\mathrm R} (p)
   \Gamma_{\mathrm {Born}}^{-1}(p) \right]_{p^2=\mu^2} = 1 \,,
\end{equation}
where $\mu$ is the renormalization scale. The renormalization function
of the fermion field ($q_R = Z_q\,q$) is given by
\be
\left.  Z_q(\mu)= \Lambda_q(p)\right|_{p^2=\mu^2} \,, \quad
  \Lambda_q(p)= \frac{ {\rm Tr} \left( - {\rm i} \sum_\lambda
   \gamma_\lambda
           \sin (a p_\lambda) a  S^{-1}(p) \right) }
           {12 \sum_\lambda \sin^2 (a p_\lambda) }
  \label{defzq}
\ee
where $\Lambda_q(p) $ is the projection of the fermion propagator onto
the tree level or Born massless quark propagator.
The renormalization factor $Z$ is calculated from the condition
\be
\left.   Z_q^{-1} \, Z\, \Lambda(p)\right|_{p^2=\mu^2} = 1 \,, \quad
   \Lambda(p)=\frac{1}{12} \,  {\rm Tr} \left[ \Gamma(p) \,
   \Gamma^{-1}_{\rm Born} (p) \right]
   \label{calcz}
\ee
where $\Lambda(p)$ is the projected amputated Green's function and
$\Gamma_{\rm Born} (p)$ is the Born term of the vertex function. 
Finally, $Z$ has to be extrapolated to the chiral limit. Note that
Eq. (\ref{calcz}) is not afflicted with $O(a)$ lattice artifacts, which
are associated with operators of opposite chirality and drop out when
the trace is taken. 

\begin{table}[b!]
\begin{center}
\begin{tabular}{|c|c|c|}\hline
$\,\,\,\kappa_\ell\,\,\,$ & $\,\,\,\kappa_s\,\,\,$ & $\,\,\,am_\pi\,\,\,$ \\ \hline
$\,\,\,0.120900\,\,\,$ & $\,\,\,0.120900\,\,\,$ &  $\,\,\,0.1757(10)\,\,\,$ \\ 
$\,\,\,0.120920\,\,\,$ & $\,\,\,0.120920\,\,\,$ & $\,\,\,0.1647(4)\phantom{0}\,\,\,$  \\ 
$\,\,\,0.120950\,\,\,$ & $\,\,\,0.120950\,\,\,$ & $\,\,\,0.1508(4)\phantom{0}\,\,\,$  \\ 
$\,\,\,0.120990\,\,\,$ & $\,\,\,0.120990\,\,\,$ & $\,\,\,0.1285(7)\phantom{0}\,\,\,$  \\ 
$\,\,\,0.121021\,\,\,$ & $\,\,\,0.121021\,\,\,$ & $\,\,\,0.1089(21)\,\,\,$ \\ \hline
\end{tabular}
\end{center}
\caption{Parameters $\kappa_l$ ($l=u,\,d$), $\kappa_s$ and pion masses of our lattice ensembles at
  $\beta=5.50$.}  
\label{tab2}
\end{table}

The calculations are done for five sets of mass-degenerate quarks,
$m_u=m_d=m_s$, on $32^3\times 64$ lattices at
$\beta=5.50$~\cite{Bietenholz:2011qq}, with the SLiNC action. The
clover parameter, $c_{\rm sw}$, was set to 2.65, and the stout
parameter, $\omega$, to 0.1~\cite{Cundy:2009yy}. The parameters of the
corresponding gauge field configurations are listed in
Table~\ref{tab2}. In terms of the hopping parameter $\kappa_q$\,, the quark masses are given by
\begin{equation}
am_q=\frac{1}{2\kappa_q}-\frac{1}{2\kappa_{0\,c}}
\end{equation}
with $\kappa_{0\,c}=0.121099(3)$. We use momentum
sources~\cite{Gockeler:1998ye} to compute the Green's functions
of Eq. (\ref{Gdef}) and discard quark-line disconnected contributions. Thus,
unless stated otherwise, our renormalization factors refer to flavor
nonsinglet operators.

To convert the renormalization factors to physical scales $\mu$,
we need to know the lattice spacing $a$. Our strategy to set the scale is to use singlet quantities~\cite{Horsley:2013wqa}, which are flat in $\delta m_q = m_q-\bar{m}$ up to corrections of $O(\delta m_u^2 +\delta m_d^2 + \delta m_s^2)$~\cite{Bietenholz:2011qq}, where $\bar{m}$ is the average quark mass ($\bar{m}=(2m_\ell +m_s)/3$ in our case). We find $a=0.074(2)\,\mbox{fm}$.
With this choice of the lattice spacing, the pion masses in
Table~\ref{tab2} range from $290$ to $465\,\mbox{MeV}$, 
allowing a controlled extrapolation to the chiral limit. The
renormalization factors are converted to the 
$\overline{\rm MS}$ scheme at $\mu=2$ GeV.

The lattice momenta are chosen according to 
\begin{equation}
\Lambda^2_{\mathrm{QCD}}\ll p^2 \le \left(\frac{\pi}{a}\right)^2 \, .
\label{ap_range}
\end{equation}
On $L^3\times T$ lattices with periodic spatial and antiperiodic temporal boundary conditions
\begin{equation}
p = \left(\frac{2\pi}{L}\,n_1, \frac{2\pi}{L}\,n_2, \frac{2\pi}{L}\,n_3, \frac{2\pi}{T}\,\left(n_4+\frac{1}{2}\right)\right)\,.
\end{equation}
To increase the number of momenta, we employ twisted boundary conditions to the quark fields, $p \rightarrow p + B$, with
\begin{equation}
B = \left(\frac{2\pi}{L}\,\theta_1, \frac{2\pi}{L}\,\theta_2, \frac{2\pi}{L}\,\theta_3, \frac{2\pi}{T}\,\theta_4 \right)\,.
\end{equation}
Thus for lattices with $T=2L$ then
\begin{equation}
p = \frac{2\pi}{L}\,\left(n_1 +\theta_1, n_2 +\theta_2, n_3 +\theta_3, \frac{1}{2}\left(n_4+\frac{1}{2} + \theta_4\right) \right)\,.
\end{equation}
We choose a fixed direction and then let $p^2$ vary. The optimal choice is along the diagonal, which leaves us with overall $O((a\,p)^2)$ corrections only, but no directional correction~\cite{Arthur:2010ht}. Our momenta and twist angles are listed in Table~\ref{tab:p2}.

\begin{table}[tbh!]
\begin{center}
\begin{tabular}{|cc|cc|cc|}
\hline
\multicolumn{2}{|c|}{$\theta=\left(0,0,0,-\frac{1}{2}\right)$} &
\multicolumn{2}{|c|}{$\theta=\left(\frac{1}{2},\frac{1}{2},\frac{1}{2},-\frac{1}{2}\right)$} &
\multicolumn{2}{|c|}{$\theta=\left(\frac{1}{4},\frac{1}{4},\frac{1}{4},0\right)$} \\
\hline
$n$ & $(a\,p)^2$ & $n$ & $(a\,p)^2$ & $n$ & $(a\,p)^2$ \\
\hline
$(1,1,1,2)$ & 0.1542 & $(0,0,0,1)$ & 0.03855 & $(0,0,0,0)$ &
0.009638\\
$(2,2,2,4)$ & 0.6169 & $(1,1,1,3)$ & 0.3470 & $(1,1,1,2)$ &
0.2410\\
$(3,3,3,6)$ & 1.3879 & $(2,2,2,5)$ & 0.9638 & $(2,2,2,4)$ &
0.7807\\
$(4,4,4,8)$ & 2.4674 & $(3,3,3,7)$ & 1.8891 & $(3,3,3,6)$ &
1.6289\\
$(5,5,5,10)$ & 3.8553 & $(4,4,4,9)$ & 3.1228 & $(4,4,4,8)$ &
2.7855\\
$(6,6,6,12)$ & 5.5517 & $(5,5,5,11)$ & 4.6649 & $(5,5,5,10)$ &
4.2505\\
$(7,7,7,14)$ & 7.5564 & $(6,6,6,13)$ & 6.5155 & $(6,6,6,12)$ &
6.0239\\
$(8,8,8,16)$ & 9.8696 & $(7,7,7,15)$ & 8.6745 & $(7,7,7,14)$ &
8.1058\\
\hline
\end{tabular}
\caption{Lattice momenta and twist angles in lattice units.}
\label{tab:p2}
\end{center}
\vspace{-0.5cm}
\end{table}
\vspace{-0.5cm}
\section{Perturbative results}
\label{sec3}

Our perturbative computation for the renormalization factors is
performed in one loop perturbation theory using a variety of fermionic
and gluonic actions. In this paper we focus on the SLiNC
action~\cite{Horsley:2008ap} providing the results for general gauge,
$\alpha$, ($\alpha=0\,(1)$ for Landau (Feynman) gauge) and action
parameters ($c_{\rm sw},\,\omega$). The Feynman diagrams entering this
computation and the procedure for their evaluation are extensively
described in Ref.~\cite{Alexandrou:2012mt}. For comparison with the
non-perturbative renormalization factors, in the results shown in this
section we employ $\omega=0.1$ for the parameter appearing in the
stout links of the fermion part, and for the clover parameter we test
both the tree-level value suggested by one-loop perturbation theory,
$c_{\rm sw} = 1$ and the value employed in the simulations, that
is~\cite{Cundy:2009yy}: $c_{\rm sw} = 2.65$. We have performed two
separate computations to one loop, as described below:
\begin{itemize}
\item[A.] up to second order in the lattice spacing, $a$, for which the
  results are given in a closed form as a function of the external
  momentum $p$ and well as $a$, $\alpha$, $c_{\rm sw}$ and $\omega$.
\item[B.] to all orders in $a$, for general values of $a$, $\alpha$,
  $c_{\rm sw}$ and $\omega$, but for specific choices for the external
  momentum; we have computed these terms for all the momenta employed
  in the non-perturbative computation. 
\end{itemize}
Our results for each computation are discussed below.

\subsection{Subtraction of ${\cal O}(g^2 a^2)$ contributions}
\label{sub31}

We compute to one loop the inverse propagator, $S^{-1}(p)$, and
the amputated two-point Green's functions of the local operators
(scalar, pseudoscalar, vector, axial-vector, tensor), $\Gamma_{\cal O}(p)$. The
$Z$-factors are extracted by applying the renormalization conditions
given in Eqs. (\ref{defzq}) - (\ref{calcz}). The tree-level values for
the fermion operators ${\cal O} = S, P, V, A$ and $T$ are
\be
\Gamma_{\rm Born}(p) = -{\rm i}\,\openone,\,\,-{\rm
  i}\,\gamma_5,\,\,-{\rm i}\,
\gamma_\mu,\,\,-{\rm i}\, \gamma_5\,\gamma_\mu,
\,\,-{\rm i}\, \gamma_5\,\smn\,,
\label{meth2}
\ee
respectively. In the mass-independent schemes which we consider, the
bare quark masses must be set to zero. For the appropriate evaluation of the $Z$-factor, the
${\cal O}(g^2\,a^2)$ terms are omitted from the Green's function
entering Eqs.~(\ref{defzq}) - (\ref{calcz}), obtaining:
\bea
\label{Zqpert}
Z_q &=&  1 + \ggcf\, \Bigl(-13.0233 +4.79201\,\alpha + c_{\rm sw} (2.01543
-4.67344\,\omega) +
1.24220 \,c_{\rm sw}^2 \nonumber \\
&&\hspace{1.9cm} +152.564 \,\omega - 541.381 \,\omega^2 - \alpha \log
(a^2\,\mu^2) \Bigr)\,, \\ [2ex]
Z_S &=&  1 + \ggcf\, \Bigl(-13.6067 -\alpha +c_{\rm sw} (18.2213 \,\omega
-6.83528) +1.36741 \,c_{\rm sw}^2  \nonumber \\
&&\hspace{1.9cm} +140.264 \,\omega -481.361 \,\omega^2 +3\log(a^2\,\mu^2)\Bigr)\,,\\ [2ex]
Z_P &=&  1 + \ggcf\, \Bigl(-21.7334  - \alpha + c_{\rm sw} (2.01543 -4.67344
\,\omega ) -1.74485\,c_{\rm sw}^2  \nonumber \\
&&\hspace{1.9cm} +201.198 \,\omega -649.867 \,\omega^2  +3\log(a^2\,\mu^2)\Bigr)\,,\\ [2ex]
Z_V &=&  1 + \ggcf\, \Bigl(-16.5029 +\frac{\alpha}{2} + c_{\rm sw} (4.2281
-10.3971 \,\omega )
+0.46414 \,c_{\rm sw}^2 \nonumber \\
&&\hspace{1.9cm} +168.263 \,\omega -584.846 \,\omega^2\Bigr)\,,\\ [2ex]
Z_A &=&  1 + \ggcf\, \Bigl(-12.5396 +\frac{\alpha}{2}  +c_{\rm sw}(1.05025
\,\omega -0.19725)
+2.02027 \,c_{\rm sw}^2 \nonumber \\
&&\hspace{1.9cm}+137.796 \,\omega -500.593 \,\omega^2  \Bigr)\,,\\ [2ex]
Z_T &=&  1 + \ggcf\, \Bigl(-13.5383 +\alpha+ c_{\rm sw} (3.49054 -8.48923
\,\omega )
+1.71918\,c_{\rm sw}^2 \nonumber \\
&&\hspace{1.9cm}+147.129 \,\omega -535.088 \,\omega^2 
 -\log(a^2\,\mu^2)\Bigr)\,,
\label{Ztpert}
\eea
where $C_F=(N^2-1)/(2N)$.

The ${\cal O}(g^2\,a^2)$ contributions of the Green's functions are
useful in non-perturbative computations of the $Z$-factors, since they
may be subtracted from the non-perturbative values; these terms are
reliable up to a limited range of the lattice spacing. 

As an example we show the expression for the projected
quark-antiquark Green's function $\Lambda_q$ defined in Eq.~(\ref{defzq}):
\be
\Lambda_q = Z_q + \Lambda_q^{(2)}
\label{LL}
\ee
using tree-level Symanzik gluons, Landau gauge $c_{\rm sw}=2.65$ and $\omega=0.1$. 
The ${\cal O}(g^2\,a^0)$ contribution gives $Z_q$,
while the ${\cal O}(g^2\,a^2)$ terms are denoted by $\Lambda_q^{(2)}$;
the latter is given by:
\be
\label{Lambda_q_2}
\Lambda_q^{(2)} = a^2\,\frac{g^2\,C_F}{16\,\pi^2}\,
\Bigg[p^2\left(-0.8825 + 0.3972\,\log (a^2\,p^2) \right)  +
  \frac{p4}{p^2}\left(1.9141 -\frac{157}{180}\,\log (a^2\,p^2) \right)  \Bigg]\,. 
\ee
Beyond ${\cal O}(a^0)$, the terms depend not only on the length, but
also on the direction of the four-vector $p$, due to the appearance of
the Lorentz noninvariant terms $p4\equiv \sum_\rho p^4_\rho$. 

\begin{figure}[!h]
{\includegraphics[scale=0.33]{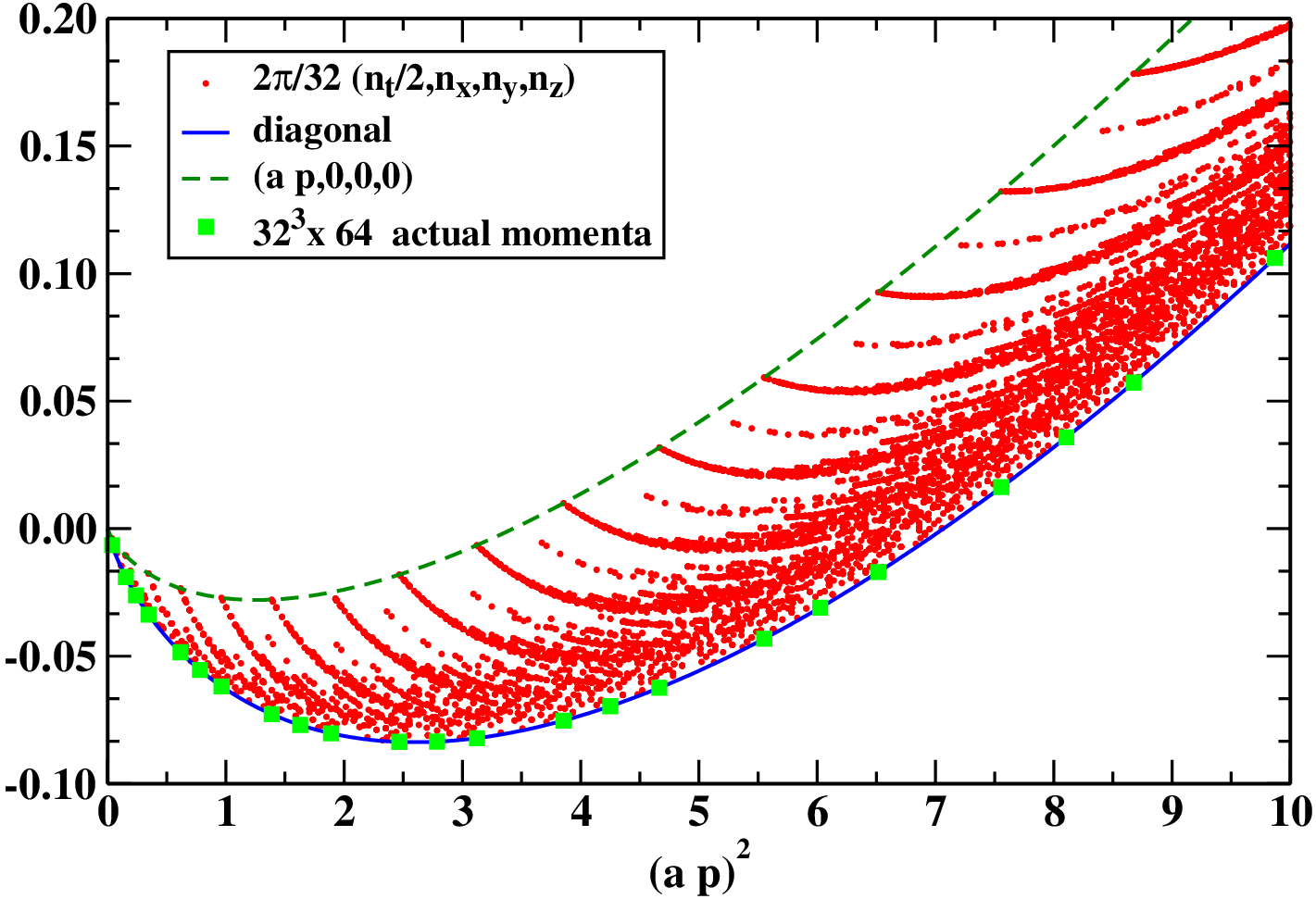}}$\quad$
{\includegraphics[scale=0.33]{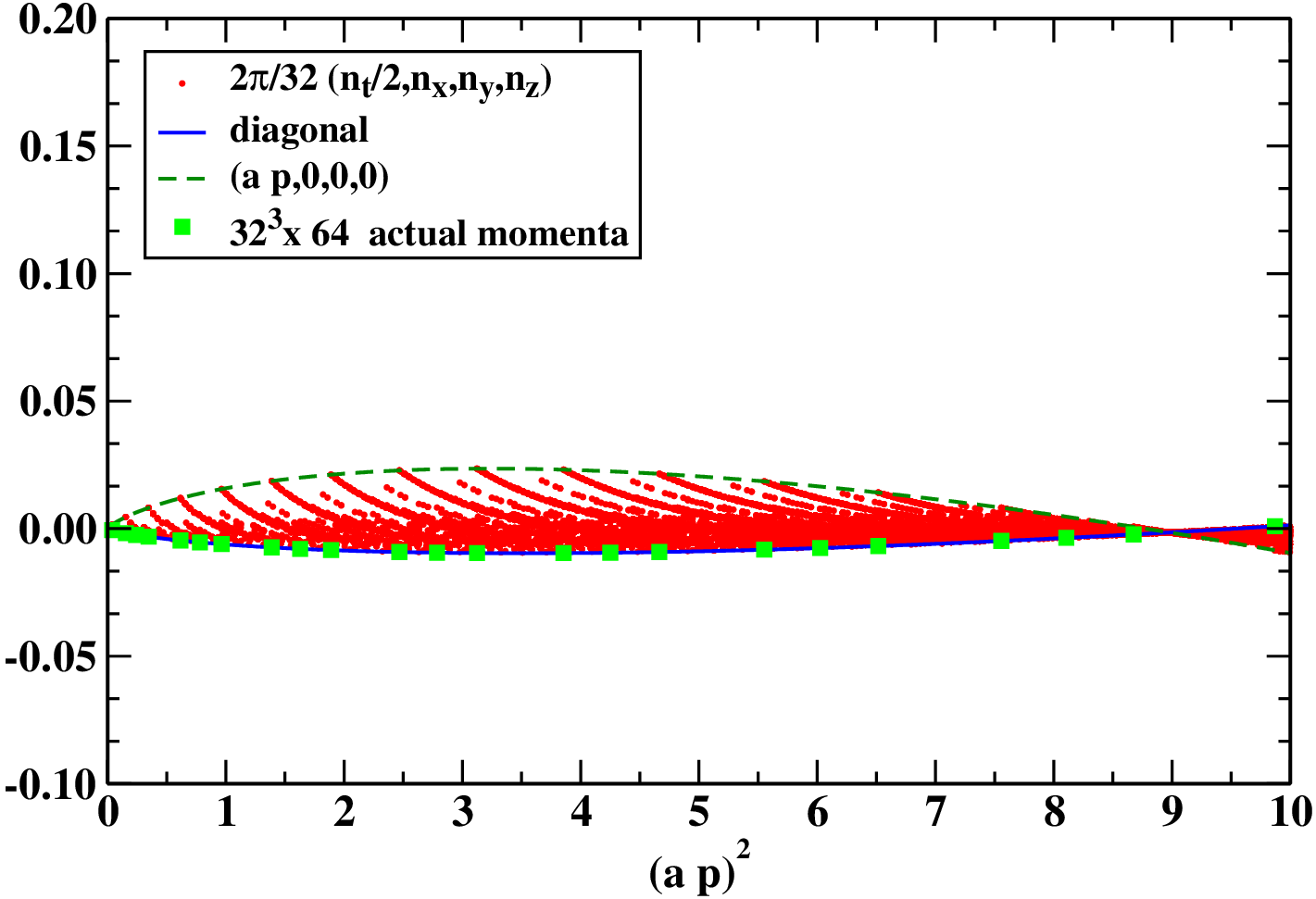}}
\vskip -0.35cm
\caption{$\Lambda_q^{(2)}$ as a function of $(a\,p)^2$ for
$\omega=0.1$ and: i) $c_{\rm sw}=2.65$ (left); ii) $c_{\rm sw}=1$
(right). For comparison, to one loop: $Z_q=1.14807$ for $c_{\rm sw}=2.65$ and
  $Z_q=0.99401$ for $c_{\rm sw}=1$.}
\label{Zq_Oa2}
\end{figure}

\begin{figure}[!h]
{\includegraphics[scale=0.33]{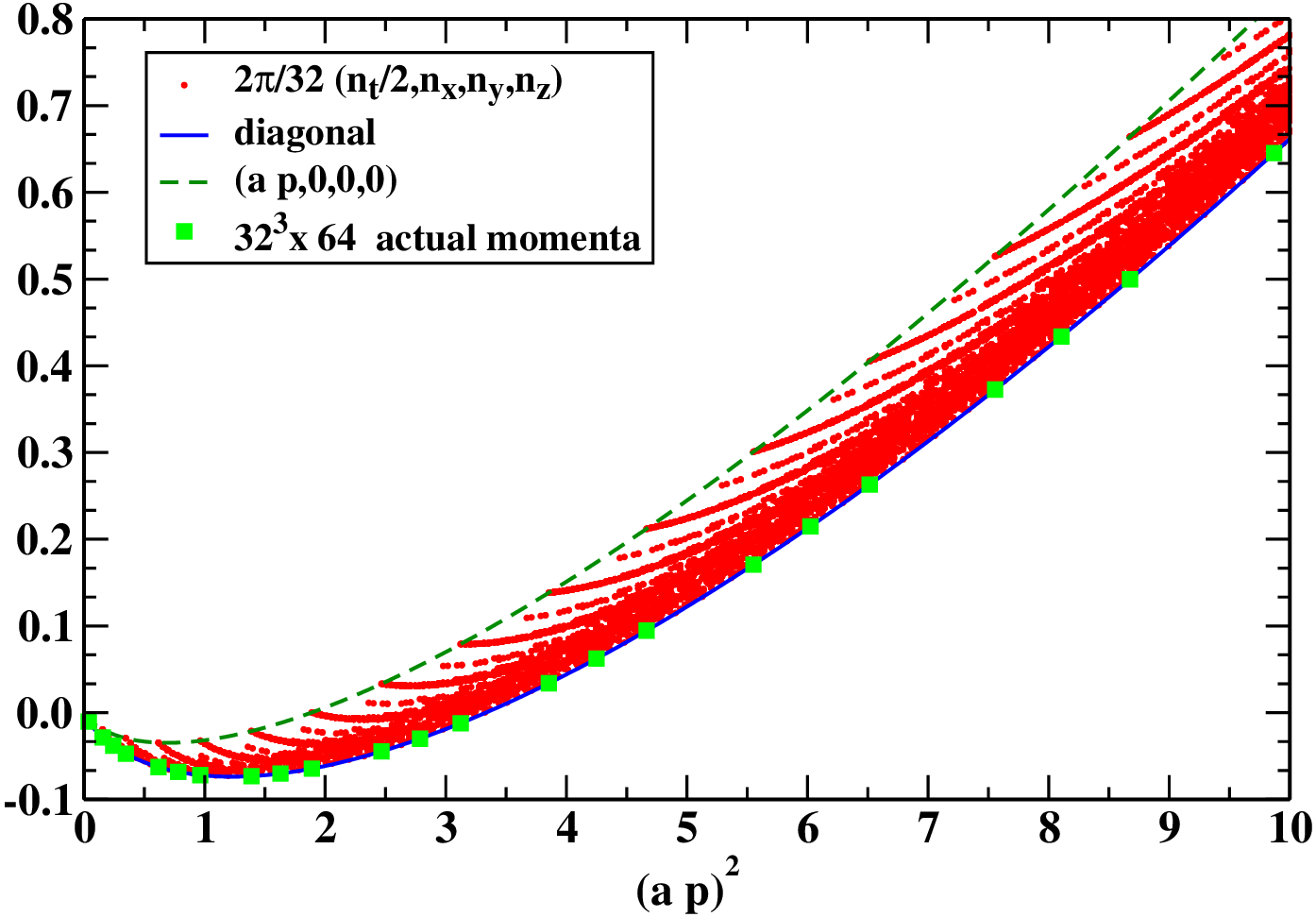}}$\quad$
{\includegraphics[scale=0.33]{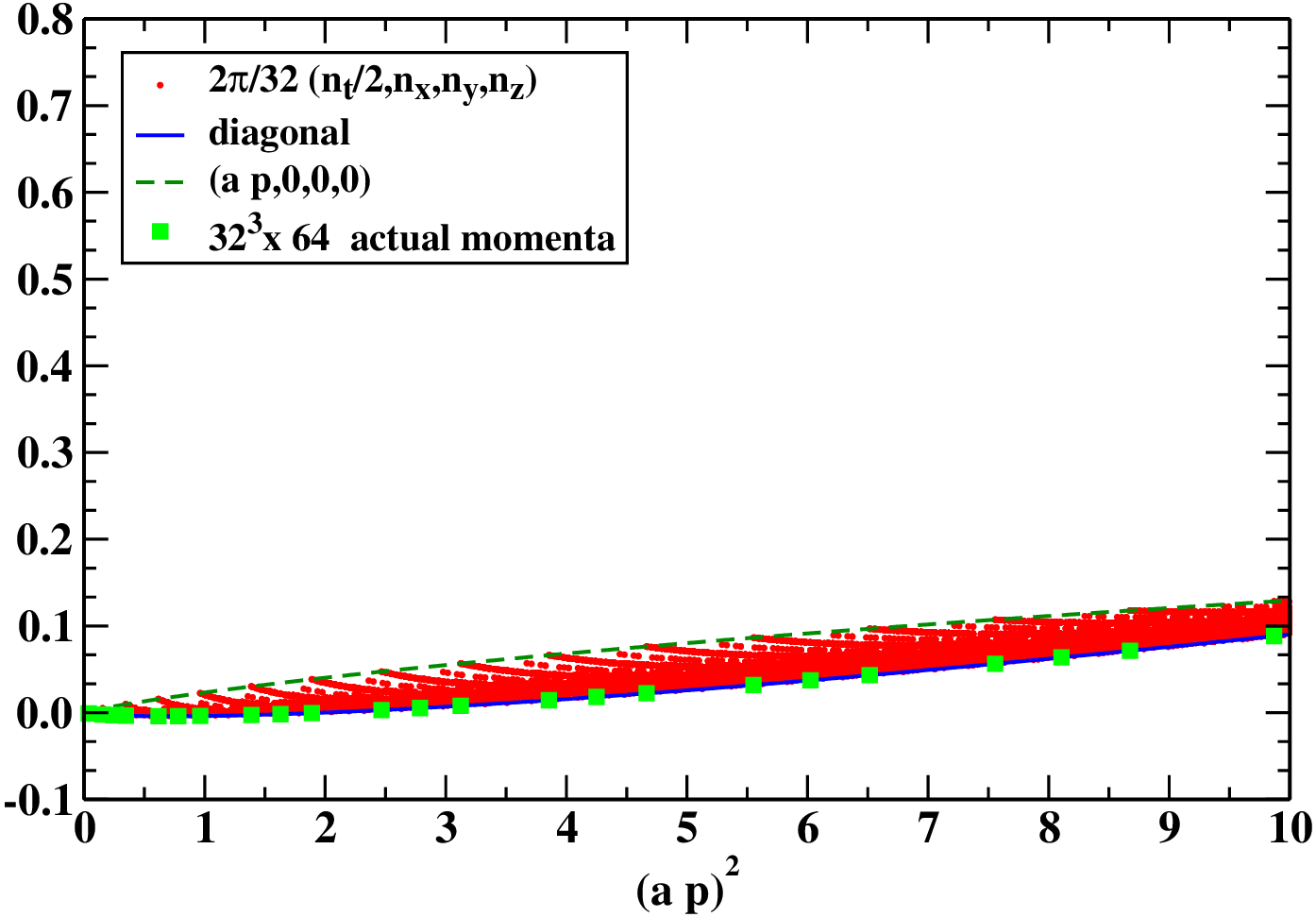}}
\vskip -0.35cm
\caption{Similar to Fig.~\ref{Zq_Oa2} for $\Lambda_A^{(2)}$. For
  comparison, to one loop: $Z_A=1.15623$ for $c_{\rm sw}=2.65$
  and $Z_A=0.97179$ for $c_{\rm sw}=1$.} 
\label{ZA_Oa2}
\end{figure}

In the left panel of Fig.~\ref{Zq_Oa2} we plot $\Lambda_q^{(2)}$ of
Eq.~(\ref{Lambda_q_2}). In the figure we highlight (in green)
the values of the momenta which we actually employ in the
non-perturbative evaluation of the $Z$-factors. An immediate observation
from the plot is that $\Lambda_q^{(2)}$ is significantly large
(especially for the diagonal momenta) as compared to the one-loop
perturbative estimate of $Z_q$ at the same action parameters
($Z_q=1.148$). This is in contrary to the case of $N_f=2$ calculated
for Wilson fermions and plaquette action, in which $c_{\rm sw}=1$, and
thus, to understand better the behavior of $\Lambda_q^{(2)}$ we plot
it in the right panel of Fig.~\ref{Zq_Oa2} for $c_{\rm sw}=1$ and
$\omega=0.1$. Indeed, the ${\cal O}(g^2\,a^2)$ terms in this case are
one order of magnitude smaller compared to the case of $c_{\rm sw}=2.65$, 
and thus under control ($Z_q=0.99401$ for $c_{\rm sw}=1$). By analogy
with Eq.~(\ref{Lambda_q_2}) and Fig.~\ref{Zq_Oa2}, we also plot in
Fig.~\ref{ZA_Oa2} the ${\cal O}(g^2\,a^2)$ terms for $Z_A$,
$\Lambda_A^{(2)}$ ($\Lambda_A \equiv Z_A + \Lambda_A^{(2)}$):
\be
\Lambda_A^{(2)} =  a^2\,\frac{g^2\,C_F}{16\,\pi^2}\,\Bigg[
  p^2\left(-0.8337 + 0.5806\,\log (a^2\,p^2) \right) 
+  \frac{p4}{p^2}\left(2.3481 -\frac{157}{180}\,\log (a^2\,p^2) \right)  \Bigg]\,.
\ee
In this case, the effect of $\Lambda_A^{(2)}$ on the one-loop
perturbative value of $Z_A{=}1.156$, is even more pronounced at 
$c_{\rm sw}{=}2.65$. Another difference between the $N_f{=}2$ and $N_f{=}3$
cases is that in the latter the diagonal momenta do not lead to the
smallest ${\cal O}(g^2\,a^2)$ effect. As a consequence, the pure
non-perturbative data (estimated with diagonal momenta) and the ${\cal
  O}(a^2)$ subtracted data have a significant numerical difference. On
the other hand, for Wilson fermions, the diagonal
momenta lead to suppressed ${\cal O}(g^2\,a^2)$ effects; we had taken
that observation on the $N_f{=}2$ case as supporting evidence for the
choice of diagonal momenta.

\FloatBarrier
\FloatBarrier

\begin{figure}[!h]
{\includegraphics[scale=0.3]{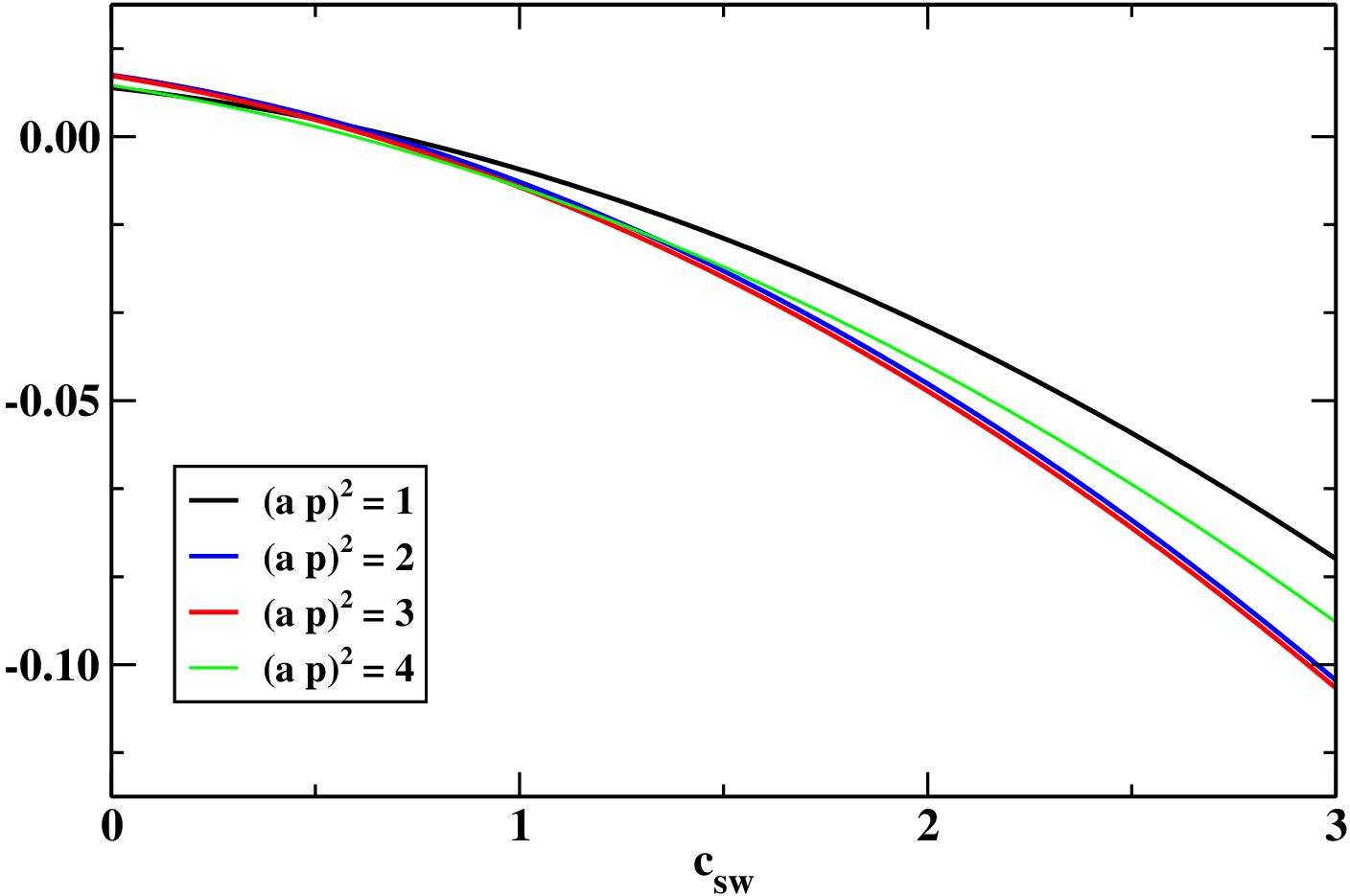}}$\quad$
{\includegraphics[scale=0.3]{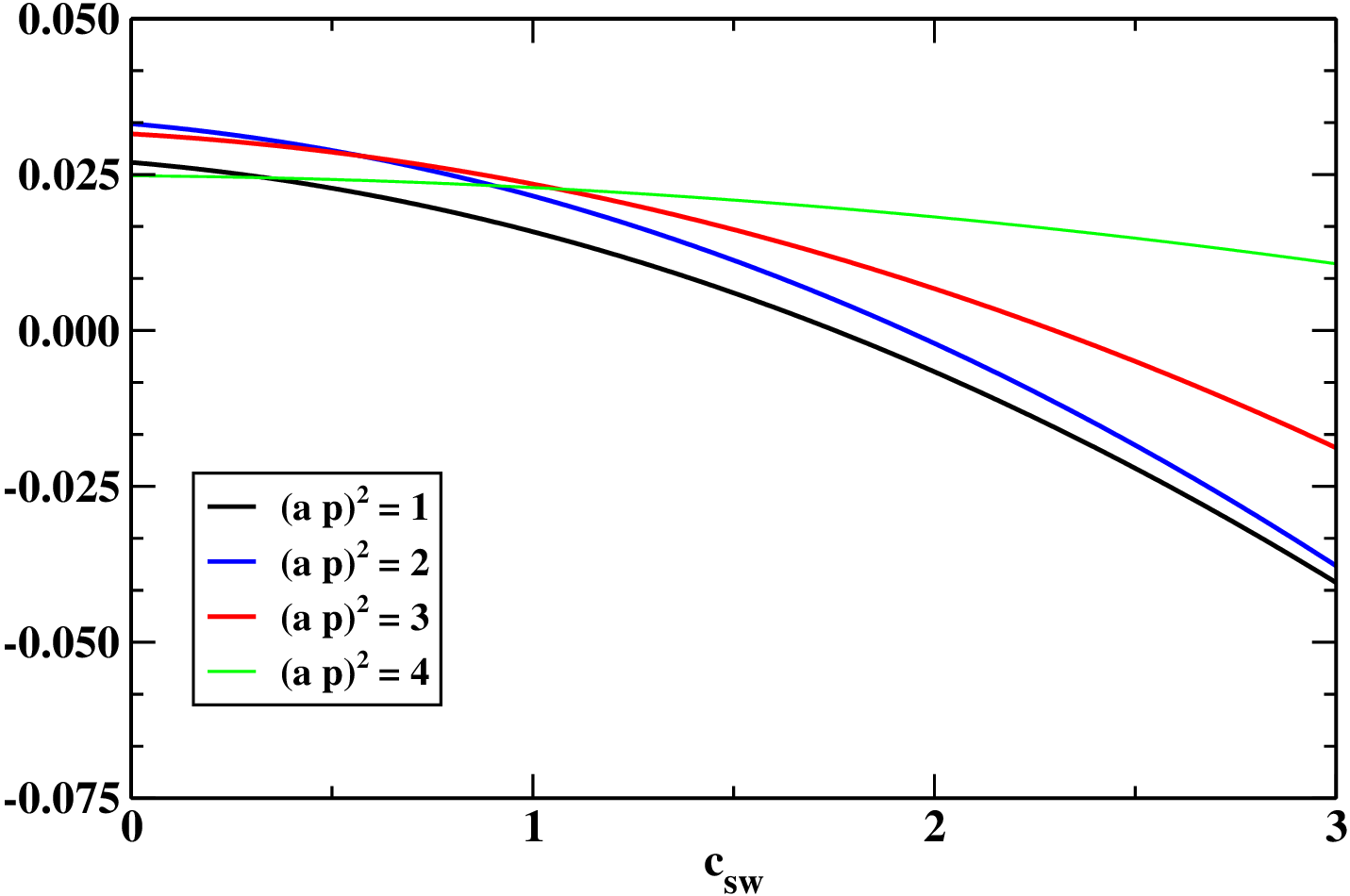}}
\vskip -0.35cm
\caption{$\Lambda_q^{(2)}$ as a function of $c_{\rm sw}$ for
  $\omega=0.1$ and for representative values of diagonal ($(a\,p)/2\,(1,1,1,1)$, left)
  and non-diagonal momenta ($(a\,p,0,0,0)$, right).}
\label{Zq_Oa2_csw}
\end{figure}

It is also interesting to investigate the $c_{\rm sw}$
dependence of $\Lambda_q^{(2)}$ for various values of $(a\,p)^2$. For
this testing we set $\omega=0.1$ and the dependence is 
shown in Fig.~\ref{Zq_Oa2_csw}. The left panel corresponds to diagonal
momenta (equal components), while the right panel to momenta with a 
nonzero component in a single direction. We observe that for $c_{\rm sw}
< 1.5$ the ${\cal O}(g^2\,a^2)$ terms are very small, while for 
$c_{\rm sw} > 2$ these contributions increase very fast.

Although we have tested both $c_{\rm sw}=1$ and $c_{\rm sw}=2.65$, we
choose to employ the tree-level value $c_{\rm sw}=1$ for consistency
to one loop perturbation theory. This will be the value used in the
subtraction of lattice artifacts from the non-perturbative estimates
(Subsec.~\ref{sec5b}).

A collection of the perturbative results for the ${\cal O}(g^2\,a^2)$
terms is given in Appendix~\ref{appA}. 

\FloatBarrier

\subsection{Complete subtraction of one-loop lattice artifacts}

\begin{figure}[!h]
{\includegraphics[scale=0.3]{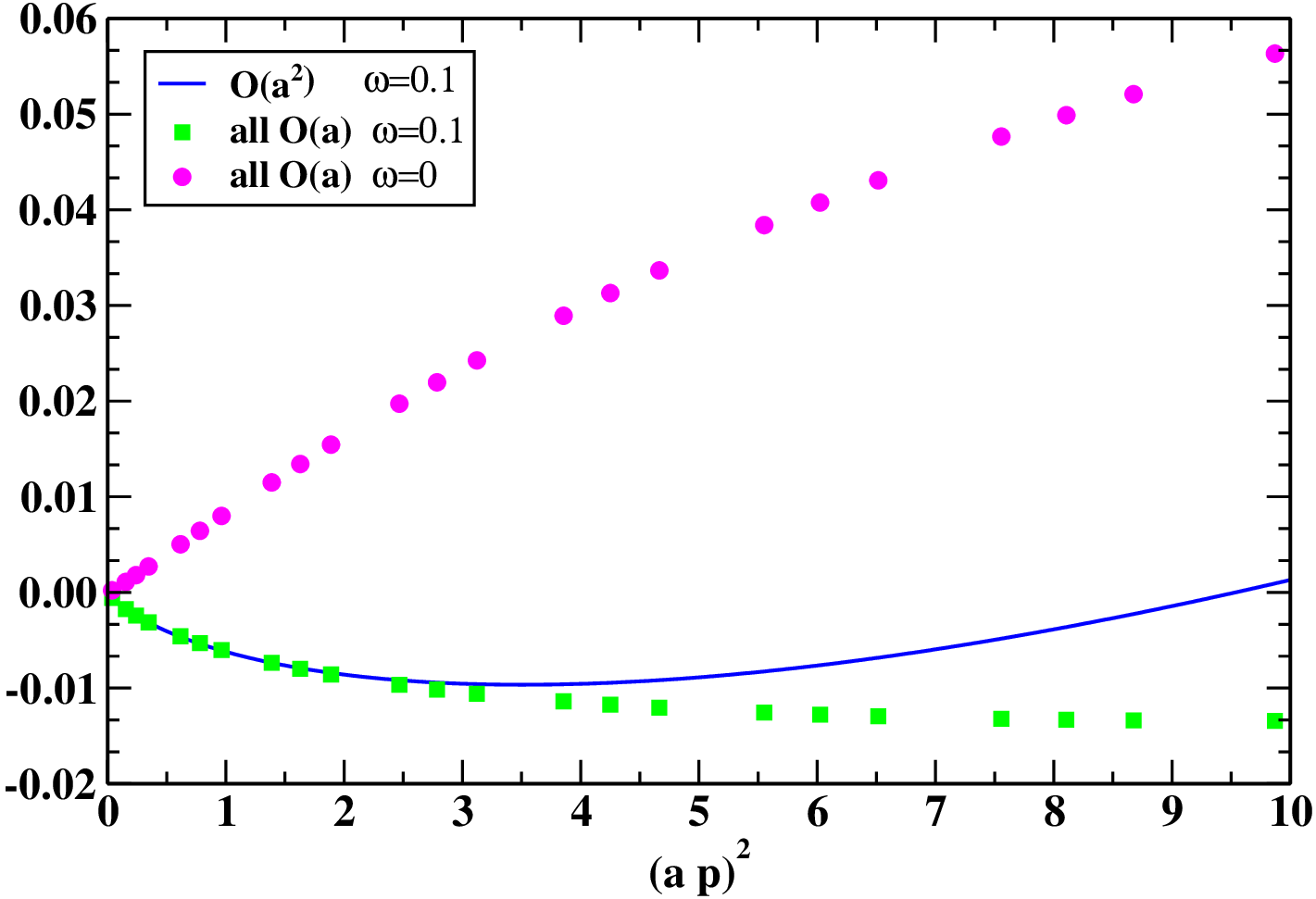}}$\qquad$
{\includegraphics[scale=0.3]{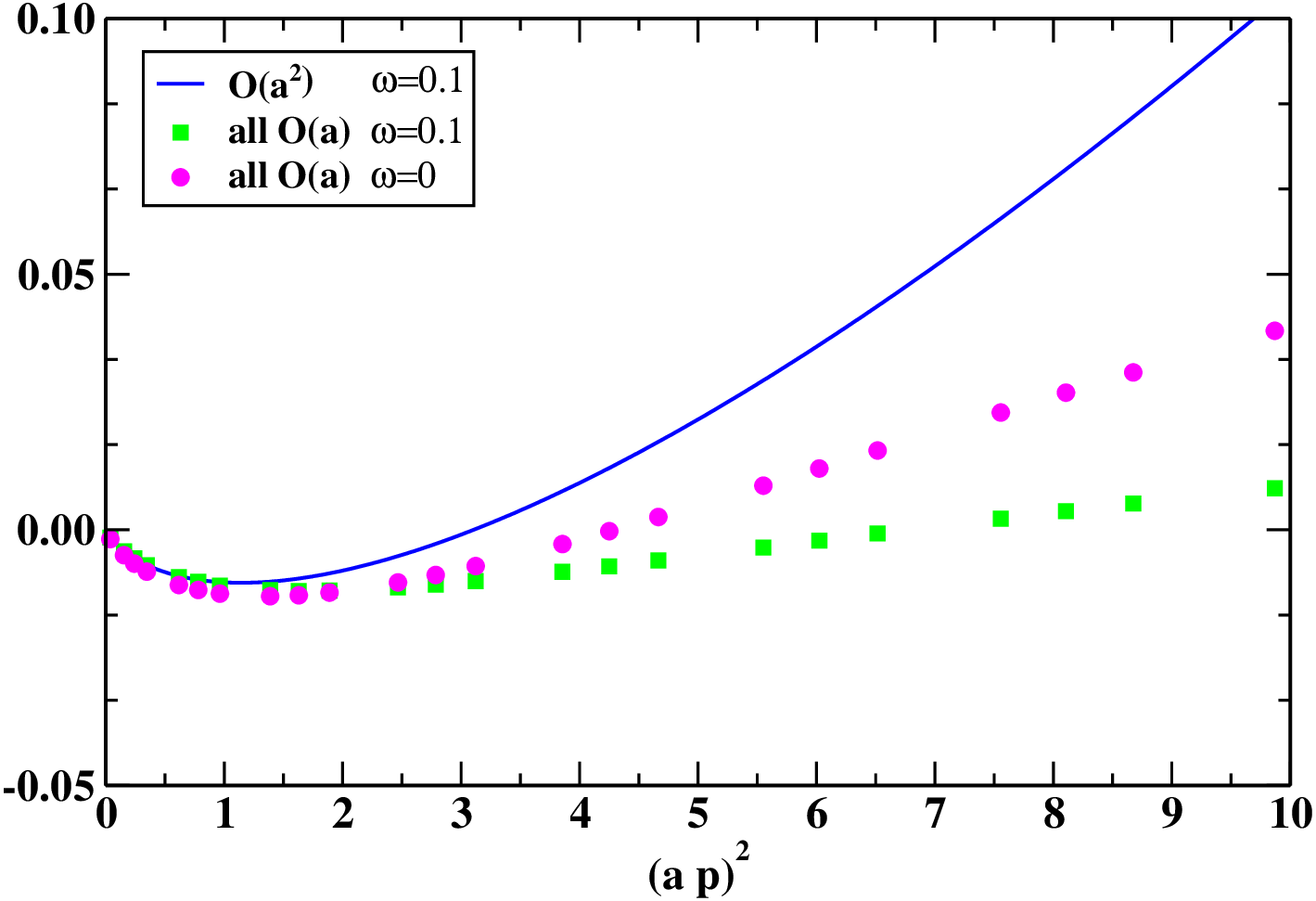}}$\qquad$
\vskip -0.35cm
\caption{Terms of all ${\cal O}(a)$ and ${\cal O}(a^2)$ for $Z_q$
(left) and $Z_T$ (right) as a function of $(a\,p)^2$.}
\label{Zq_Zt_all_Oa2}
\end{figure}

\begin{figure}[!h]
{\includegraphics[scale=0.3]{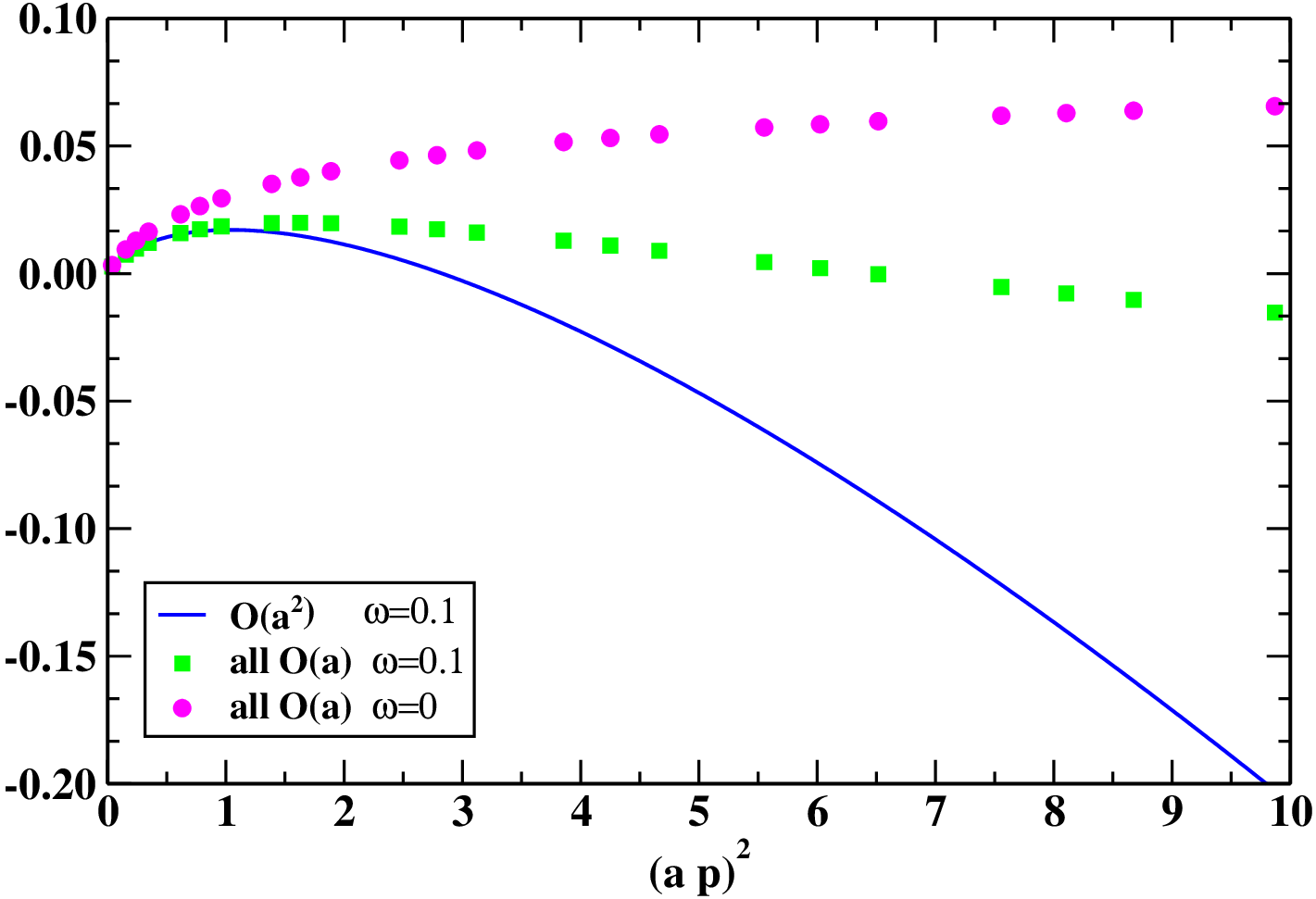}}$\qquad$
{\includegraphics[scale=0.3]{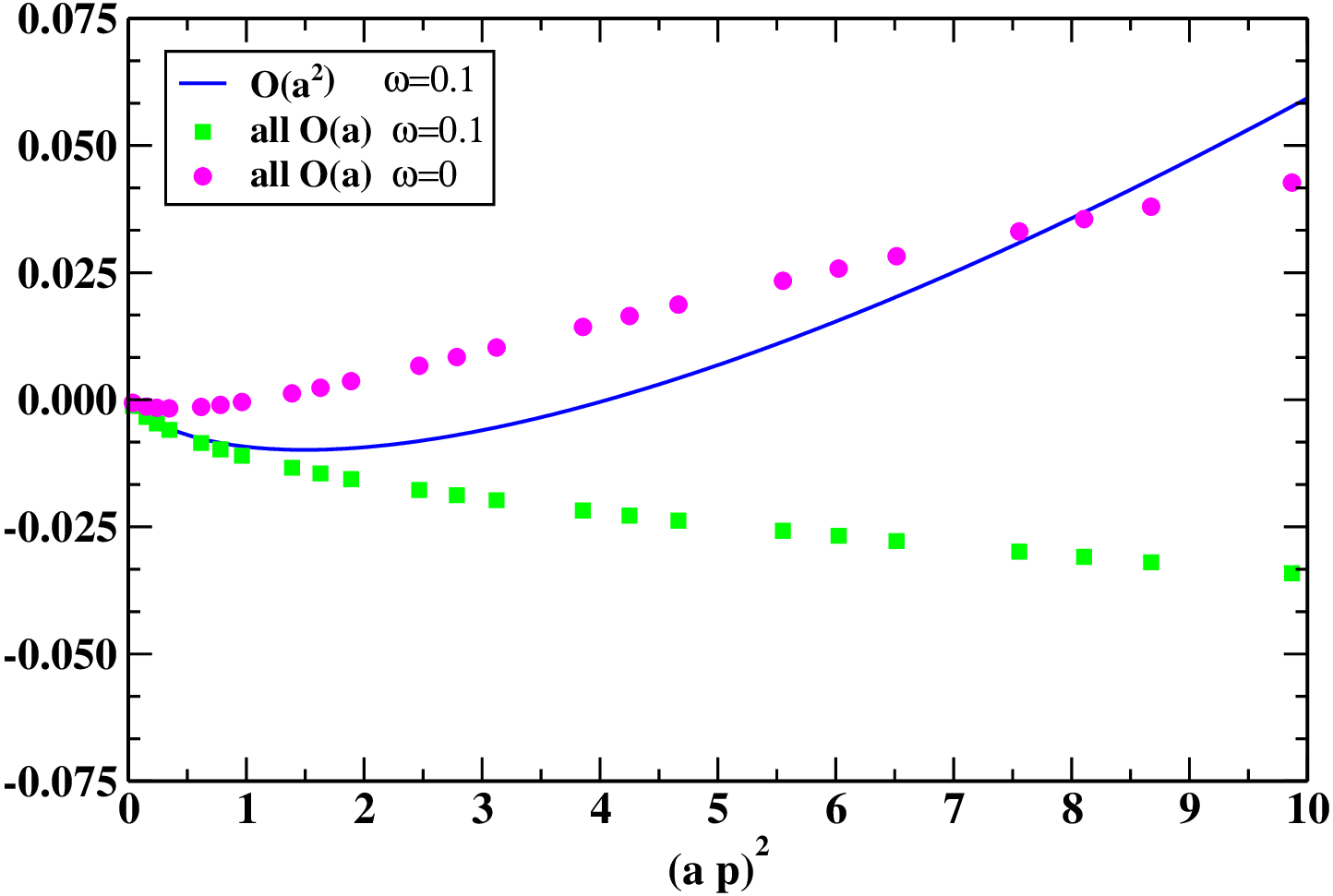}}$\qquad$
\vskip -0.35cm
\caption{Similar to Fig.~\ref{Zq_Zt_all_Oa2} for $Z_S$ (left) and $Z_P$ (right).}
\label{Zs_Zp_all_Oa2}
\end{figure}

\begin{figure}[!h]
{\includegraphics[scale=0.3]{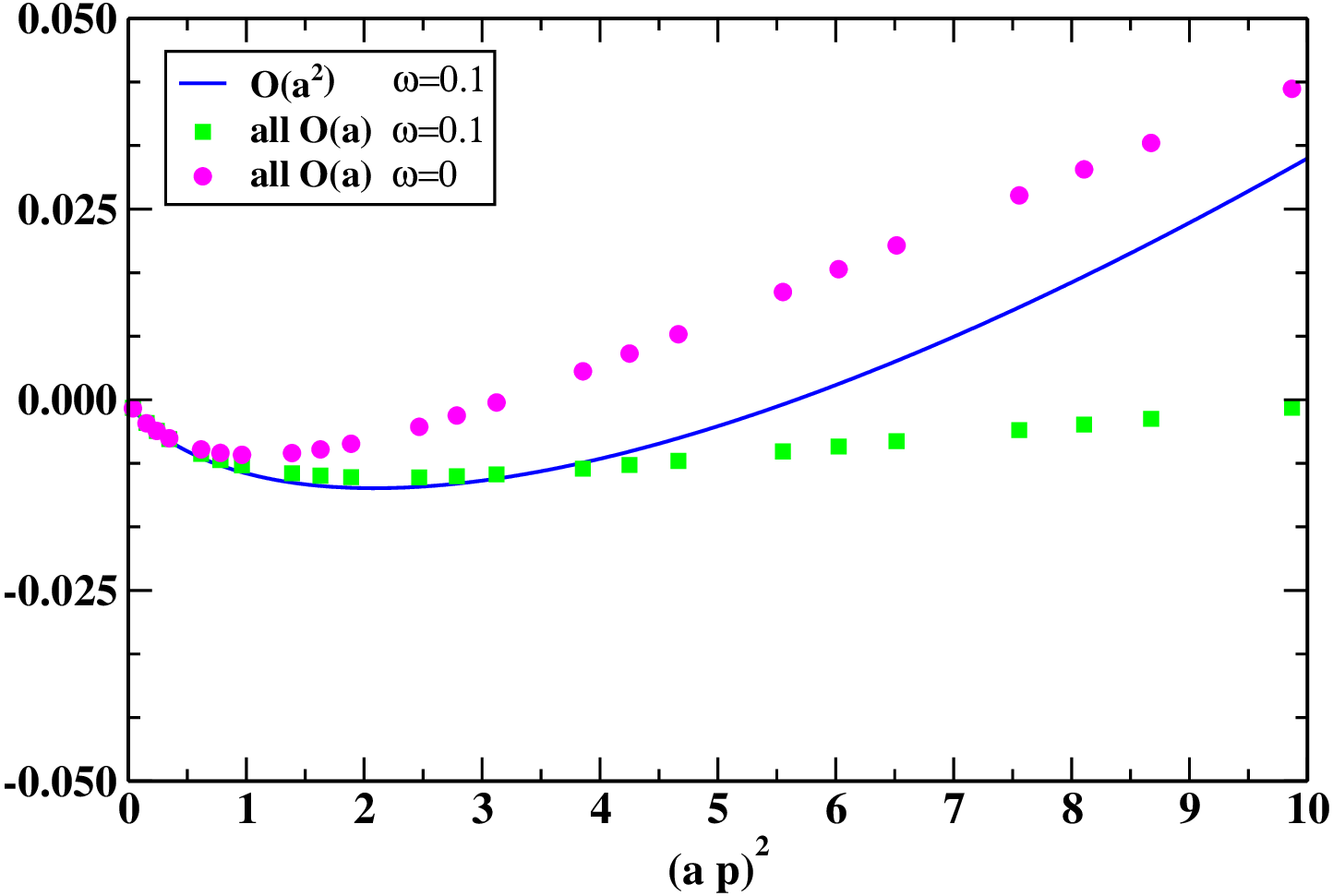}}$\qquad$
{\includegraphics[scale=0.3]{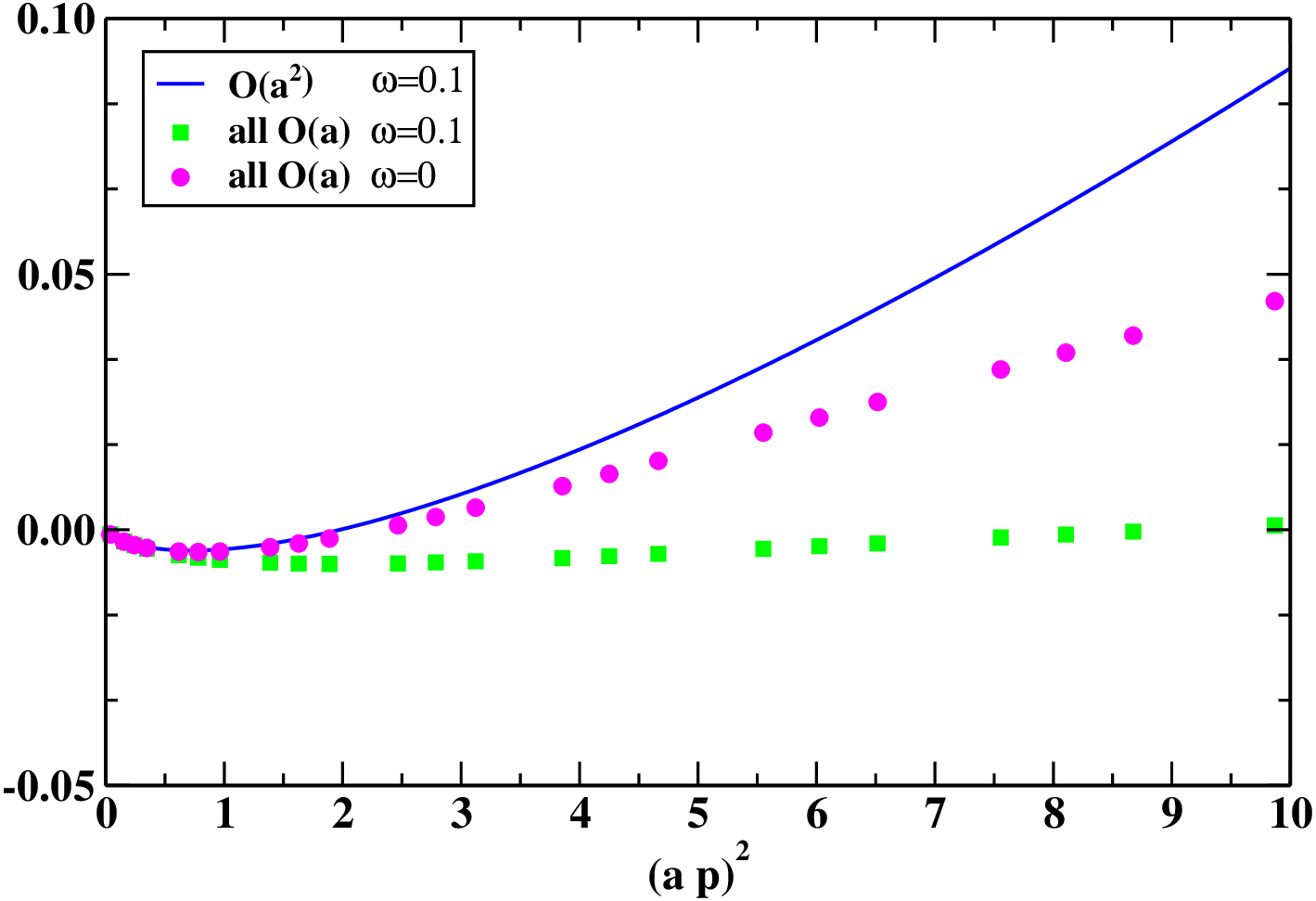}}$\qquad$
\vskip -0.35cm
\caption{Similar to Fig.~\ref{Zq_Zt_all_Oa2} for $Z_V$ (left) and
  $Z_A$ (right).}
\label{Zv_Za_all_Oa2}
\end{figure}

Here we present our results for the one-loop computation including all
orders in the lattice spacing. The main motivation for such a
calculation comes from the observation that at high values of
$(a\,p)^2$ the ${\cal O}(g^2\,a^2)$ terms are no longer under control
and become large. This does not necessarily indicate large artifacts,
since there might be cancellations with higher order artifacts. We
illustrate that this is indeed the case in this work,
demonstrating the possible complicated dependence of observables on the
lattice spacing and showing the intrinsic ${\cal O}(a)$ improvement of
the SLiNC action. Since the ${\cal O}(g^2\,a^2)$ artifacts cannot be
trustworthy for the whole range of employed momenta, it is unnatural
to subtract them from the non-perturbative estimates. Due to the nature of the
computation to all orders in $a$, the dependence on the external
momentum cannot be given in a closed form since it is included in the
propagators. Thus, we compute the one-loop expression separately for
each value of the external momentum used in the simulations. From the
latter expression one must omit the ${\cal O}(a^0)$ contributions;
this is achieved by subtracting the ${\cal O}(g^2\,a^0)$ terms,
computed analytically in Subsection~\ref{sub31} (Eqs.~(\ref{Zqpert}) -
(\ref{Ztpert})). It is interesting to compare the ${\cal O}(a^2)$
terms and the total one-loop lattice artifacts, and thus we plot the
two contributions in Figs.~\ref{Zq_Zt_all_Oa2} - \ref{Zv_Za_all_Oa2}
for $c_{\rm sw}=1$, and for two values of the stout parameter,
$\omega=0,\,0.1$. To match our non-perturbative computation we choose
diagonal momenta. A comparison of the lattice artifacts to all orders
in $a$ for $\omega=0$ and $\omega=0.1$ confirms that the presence of
the stout parameter suppresses them. Concentrating on  $\omega=0.1$,
one observes that for small values of $(a\,p)^2$ the two computations
(blue line, green points) lead to compatible results, as expected. The
${\cal O}(a^2)$ contribution increases with $(a\,p)^2$, while the
complete one-loop artifacts are much smaller (except for
$Z_q$). Nevertheless, both results are relatively small compared to
the $N_f=2$ case presented in Ref.~\cite{Constantinou:2013ada}, and
this is presumably due to the improvement of the action using the
stout smearing. This is an indication of suppressed lattice
artifacts.
\FloatBarrier
\FloatBarrier
\FloatBarrier

\section{Non-Perturbative results}
\label{sec4}

We shall present our raw results now, including the extrapolation to the chiral limit, 
the conversion to the ${\overline{\rm MS}}$ scheme and subtraction of the pion pole in case of 
the pseudoscalar density.
 
\vskip -0.3cm
\subsection{RI$'$-MOM and chiral extrapolation}
\vskip -0.3cm

\begin{figure}[!h]
{\includegraphics[scale=0.41]{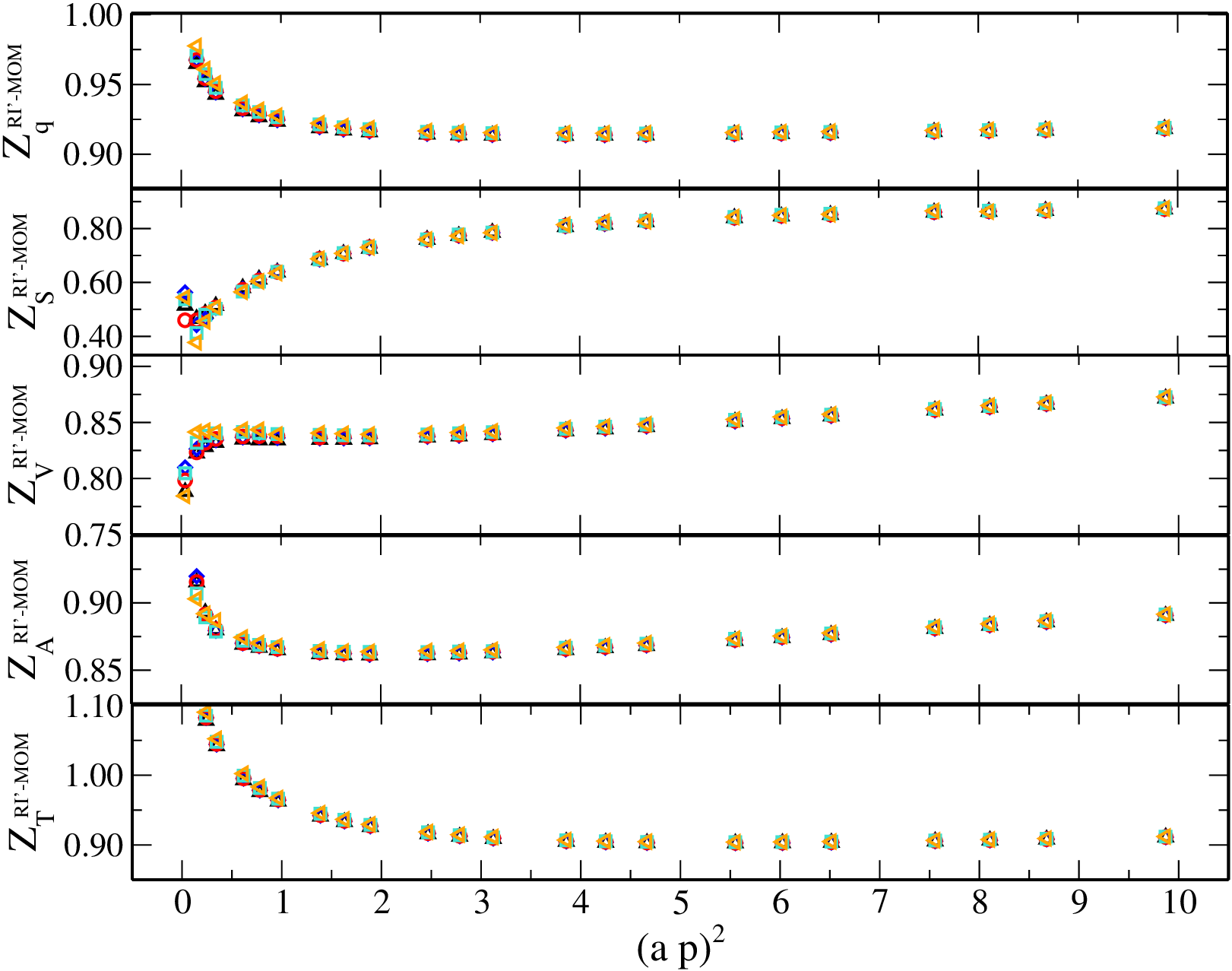}}
\caption{Results on the $Z$-factors for all pion masses versus the
  renormalization scale in lattice units. Different colors and shapes
  denote different ensembles (see also legend of Fig.~\ref{Z_vs_mpi}):
  black up triangles, blue diamonds, red circles, green squares and
  yellow left triangles stand for $m_\pi=465,\,439,\,402,\,345$, and $290$ MeV,
  respectively.}
\label{Z_vs_mpi_2}
\end{figure}

\begin{figure}[!h]
{\includegraphics[scale=0.41]{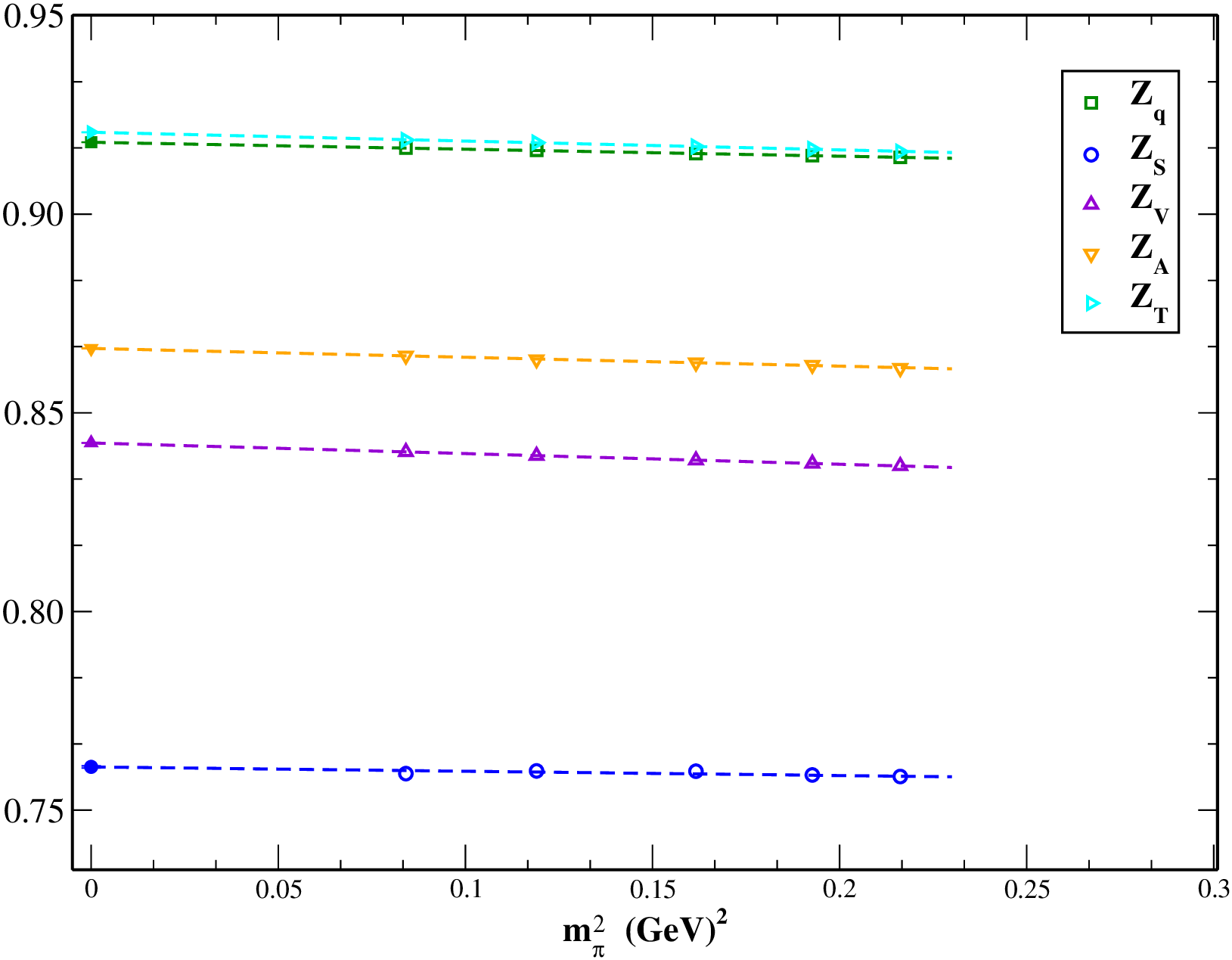}}
\vspace{-0.35cm}
\caption{The dependence of the $Z$-factors on the pion mass at $(a\,p)^2 \sim 2.5$, together with a linear fit.}
\label{Z_vs_mpi}
\end{figure}

In Fig.~\ref{Z_vs_mpi_2} we show $Z_q, Z_S, Z_V, Z_A$ and $Z_T$ as a function of the renormalization 
scale for our various pion masses. The dependence on the pion mass is found to be very weak, 
as shown in Fig.~\ref{Z_vs_mpi} for a particular value of the scale, and well represented by 
a linear curve. In the following we extrapolate the $Z$-factors linearly to the chiral limit 
for each value of $(ap)^2$. The $Z$-factor of the pseudoscalar density, $Z_P$, needs to be treated 
separately, because it suffers from the pion pole in the Green's function of Eq. (\ref{Gdef}). 
This we will deal with in Sec.~\ref{subsec43}.
\FloatBarrier

\FloatBarrier


\vspace{-0.7cm}
\subsection{Conversion to $\overline{\rm{MS}}$}
\label{sub42}

\vspace{-0.3cm}
Although RI$'$-MOM is a convenient scheme to compute the
renormalization factors, our aim is to present them in the
$\overline{\rm MS}$ scheme. Starting from the results in RI$'$-MOM at
various scales, $\mu$, we convert them to the $\overline{\rm MS}$
scheme at a reference scale; we will set that scale to 2 GeV. 
The conversion factors do not depend on the regularization scheme
(and, thus, they are independent of the lattice discretization) when
expressed in terms of the renormalized coupling constant. However,
expressing them in terms of the bare coupling constant introduces a
dependence on the action. The conversion factors, 
$C_{\cal O}^{\mbox{\scriptsize RI$^{\prime}$-MOM},\,{\overline{\rm MS}}}$, 
and the expression for running the
scale to 2 GeV in $\overline{\rm MS}$, $R(\mu,2\,{\rm GeV})$, are defined such that:
\be
Z_{\cal O}^{\overline{\rm MS}}(2\,{\rm GeV}) = R(\mu,2\,{\rm GeV})\, C_{\cal
  O}^{\mbox{\scriptsize RI$^{\prime}$-MOM},\,{\overline{\rm MS}}}\, Z_{\cal O}^{\mbox{\scriptsize RI$^{\prime}$-MOM}}(\mu) \,,
\ee
and their perturbative expressions are available up to three loops. For
the relation between the renormalized coupling constant, $g_R$, and the
bare one, $g$: $g_R{=}Z^{-1}_g\,g$, we use the two-loop results of
Ref.~\cite{Bode:2001uz} for $Z_g$ corresponding to clover fermions and
Wilson gluons. $R(\mu,2\,{\rm GeV})$ is expressed in terms of
$\Lambda_{\overline{\rm MS}}$ which for $N_f{=}3$ was estimated to be
339 MeV~\cite{PDG}. The conversion factors we use are adapted from
Ref.~\cite{Gracey:2003yr} and are applicable to the naive dimensional
regularization (NDR) of the $\overline{\rm MS}$ scheme~\cite{Buras:1989xd}, 
in which $C_P{=}C_S$. Moreover, the conversion factor from the RI$'$-MOM
to the ${\overline{\rm MS}}$ scheme for the vector and axial-vector
operators is 1. 

\begin{figure}[!h]
{\includegraphics[scale=0.36]{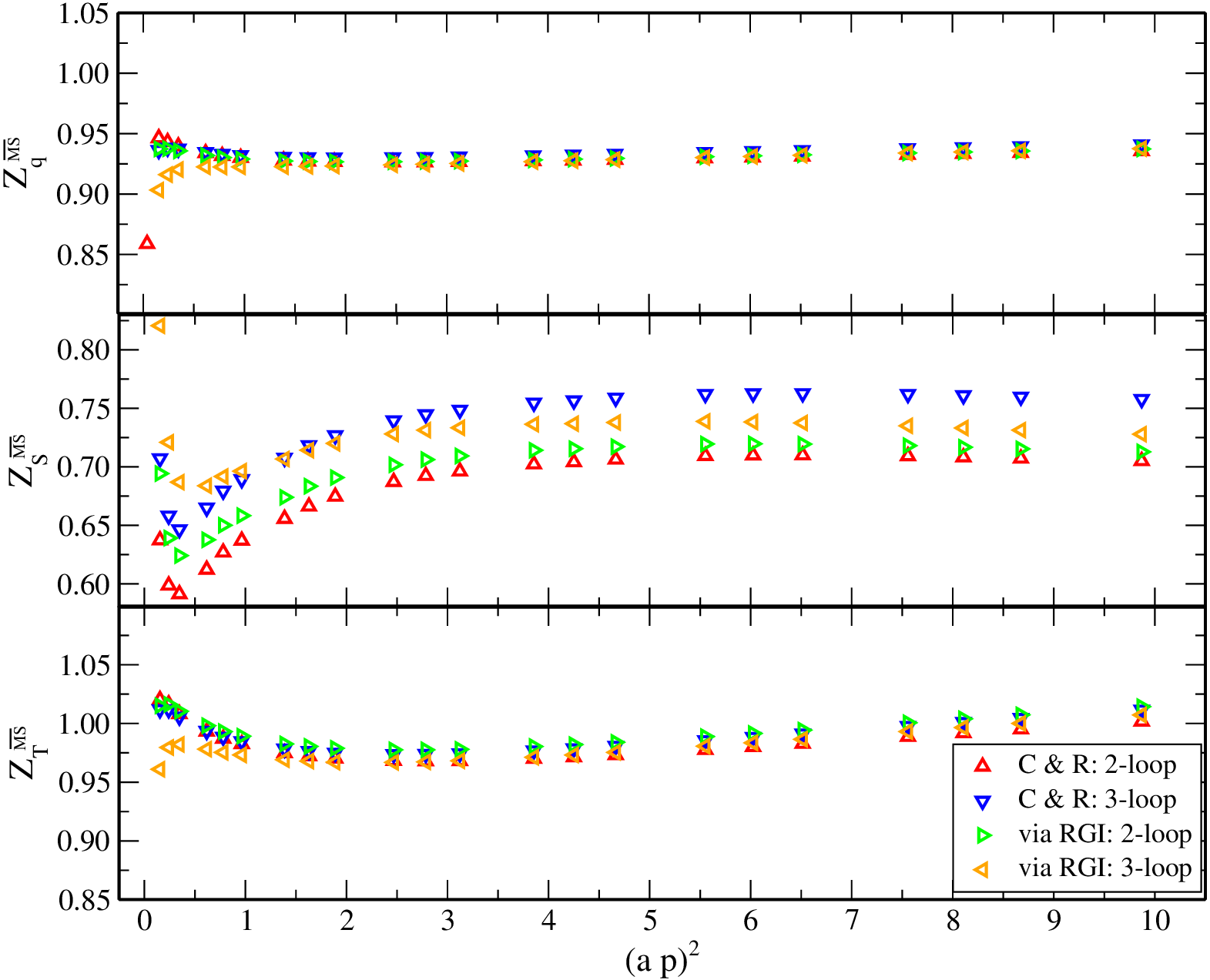}}
\vspace{-0.75cm}
\caption{$Z_q^{\overline{\rm MS}},Z_S^{\overline{\rm MS}},
Z_T^{\overline{\rm MS}}$ ($m_\pi{=}465$ MeV) using two-loop (red up
  triangles) and three-loop (blue down triangles) expressions for
  $C_{\cal O}^{\mbox{\scriptsize RI$^{\prime}$-MOM},\,{\overline{\rm MS}}}\,R(\mu,2\,{\rm
    GeV})$. Green right triangles (yellow left triangles) correspond
  to the two- (three-) loop expressions of Eq.~(\ref{DZ_RGI}) using
  the intermediate RGI scheme.}
\label{Zq_Zs_Zt_CR}
\end{figure}

In an alternative procedure the RGI scheme is used as an intermediate
scheme to obtain the conversion factors for the operators. Those
factors $C^{\mbox{\scriptsize RI$^{\prime}$-MOM},\,{\overline{\rm MS}}}$ are found from relating the
renormalization function in $\overline{\rm MS}$ (at 2 GeV) and
RI$'$-MOM (at a scale $\mu$):
\be
Z^{\rm RGI}_{\cal O} = Z_{\cal O}^{\overline{\rm MS}} (2\,{\rm GeV})
\,\Delta Z_{\cal O}^{\overline{\rm MS}}
(2\,{\rm GeV}) = Z_{\cal O}^{\mbox{\scriptsize RI$^{\prime}$-MOM}} (\mu) \,
\Delta Z_{\cal O}^{\mbox{\scriptsize RI$^{\prime}$-MOM}}(\mu)\,,
\ee
and thus
\begin{eqnarray}
\label{Z_RGI}
&Z_{\cal O}^{\overline{\rm MS}} (2\,{\rm GeV})  =   
C_{{\cal O},\rm RGI}^{\mbox{\scriptsize RI$^{\prime}$-MOM},\,
{\overline{\rm MS}}} (\mu,2\,{\rm GeV})\,Z_{\cal O}^{\mbox{\scriptsize RI$^{\prime}$-MOM}} (\mu)\,,&\nonumber
\\
&C_{{\cal O},\rm RGI}^{\mbox{\scriptsize RI$^{\prime}$-MOM},\,{\overline{\rm MS}}} (\mu,2\,{\rm GeV}) = \frac{\Delta
  Z_{\cal O}^{\mbox{\scriptsize RI$^{\prime}$-MOM}}(\mu)}{\Delta Z_{\cal O}^{\overline{\rm MS}}(2\,{\rm  GeV})}\,.&
\end{eqnarray}
The quantity $\Delta Z_{\cal O}^{\mathcal S}(\mu)$ in scheme $S$ is expressed in
terms of the $\beta-$function and the anomalous dimension of the
operator under study, $\gamma_{\cal O}^S \equiv \gamma^S$:
\be
\Delta Z_{\cal O}^{\mathcal S} (\mu) =
  \left( 2 \beta_0 \frac {{g^{\mathcal S} (\mu)}^2}{16 \pi^2}\right)
^{-\frac{\gamma_0}{2 \beta_0}}
 \exp \left \{ \int_0^{g^{\mathcal S} (\mu)} \! \mathrm d g'
  \left( \frac{\gamma^{\mathcal S}(g')}{\beta^{\mathcal S} (g')}
   + \frac{\gamma_0}{\beta_0 \, g'} \right) \right \}\,.
\ee
To three-loop approximation $\Delta Z_{\cal O}^{\mathcal S} (\mu)$ takes the form:
\bea
\label{DZ_RGI}
\Delta Z_{\cal O}^{\mathcal S} (\mu) &\hspace{-0.1cm}=\hspace{-0.1cm}&
  \left( 2 \beta_0 \frac {{g^{\mathcal S} (\mu)}^2}{16 \pi^2}\right)
^{-\frac{\gamma_0}{2 \beta_0}}
\Bigg( 1  + \frac {g^{\mathcal S} (\mu)^2}{16 \pi^2}\,
\frac{\beta_1 \gamma_0-\beta_0 \gamma^S_1}{2 \beta_0^2} +  \\
&& 
\frac {{g^{\mathcal S} (\mu)}^4}{(16 \pi^2)^2}\,
   \frac{-2 \beta_0^3 \gamma^S_2+\beta_0^2 (\gamma^S_1 (2
    \beta_1+\gamma^S_1)+2 \beta_2 \gamma_0)-2 \beta_0 \beta_1
    \gamma_0 (\beta_1+\gamma^S_1)+\beta_1^2 \gamma_0^2}{8
     \beta_0^4}  \Bigg)\,. \nonumber
\eea
The expressions for the coupling $g^S(\mu)$ in the ${\overline{\rm MS}}$ 
and in the RI$'$-MOM schemes coincide to three loops and read~\cite{Alekseev:2002zn}:
\bea
\label{gS}
\frac{{g^S(\mu)}^2}{16\pi^2} =
\frac{1}{\beta_0\, L} 
 - \frac{\beta_1}{\beta_0^3} 
\frac{\log L}
{L^2} 
+\frac{1}{\beta_0^5}
\frac{\beta_1^2 \log^2 L
- \beta_1^2 \log L  +
\beta_2 \beta_0 - \beta_1^2}{L^3} \,,\quad L = \log \frac{\mu^2}{\Lambda^2_{\overline{\rm MS}}}\,.
\eea
In Appendix~\ref{appB} we give the definitions of the
$\beta-$function and the anomalous dimension for the fermion field and
local operators, as well as their perturbative coefficients to three loops.
In Fig.~\ref{Zq_Zs_Zt_CR} we demonstrate the effects in the $Z$-factors
resulting from the use of the two- and three-loop expressions for $C_{\cal O}^{\mbox{\scriptsize RI$^{\prime}$-MOM},\,
{\overline{\rm MS}}}$ and $R(\mu,2\,{\rm GeV})$, and the corresponding
expressions for the alternative conversion (via RGI) as given by
Eq.~(\ref{DZ_RGI}). The ensemble used for this Figure corresponds to
$m_\pi=465$ MeV. We find that the discrepancies are at $8\%$
maximum for $Z_S$ (and consequently $Z_P$ discussed later). For the standard
conversion factors $C_{\cal O}^{\mbox{\scriptsize RI$^{\prime}$-MOM},\,{\overline{\rm MS}}}$ the
difference in the two-loop values of $Z_g$ between the Wilson and
tree-level Symanzik gluons is expected to be (based on their one-loop
difference) within this systematic error. The conversion via the RGI
scheme has the property that it uses continuum results; in
the rest of the paper we use the intermediate RGI scheme and employ
the three-loop result of Eq.~(\ref{DZ_RGI}) for all the
conversions. In the final results presented in Sec.~\ref{sec5} we
also give a systematic error due to differences in the conversion
factor.
\FloatBarrier

\begin{figure}[!h]
{\includegraphics[scale=0.38]{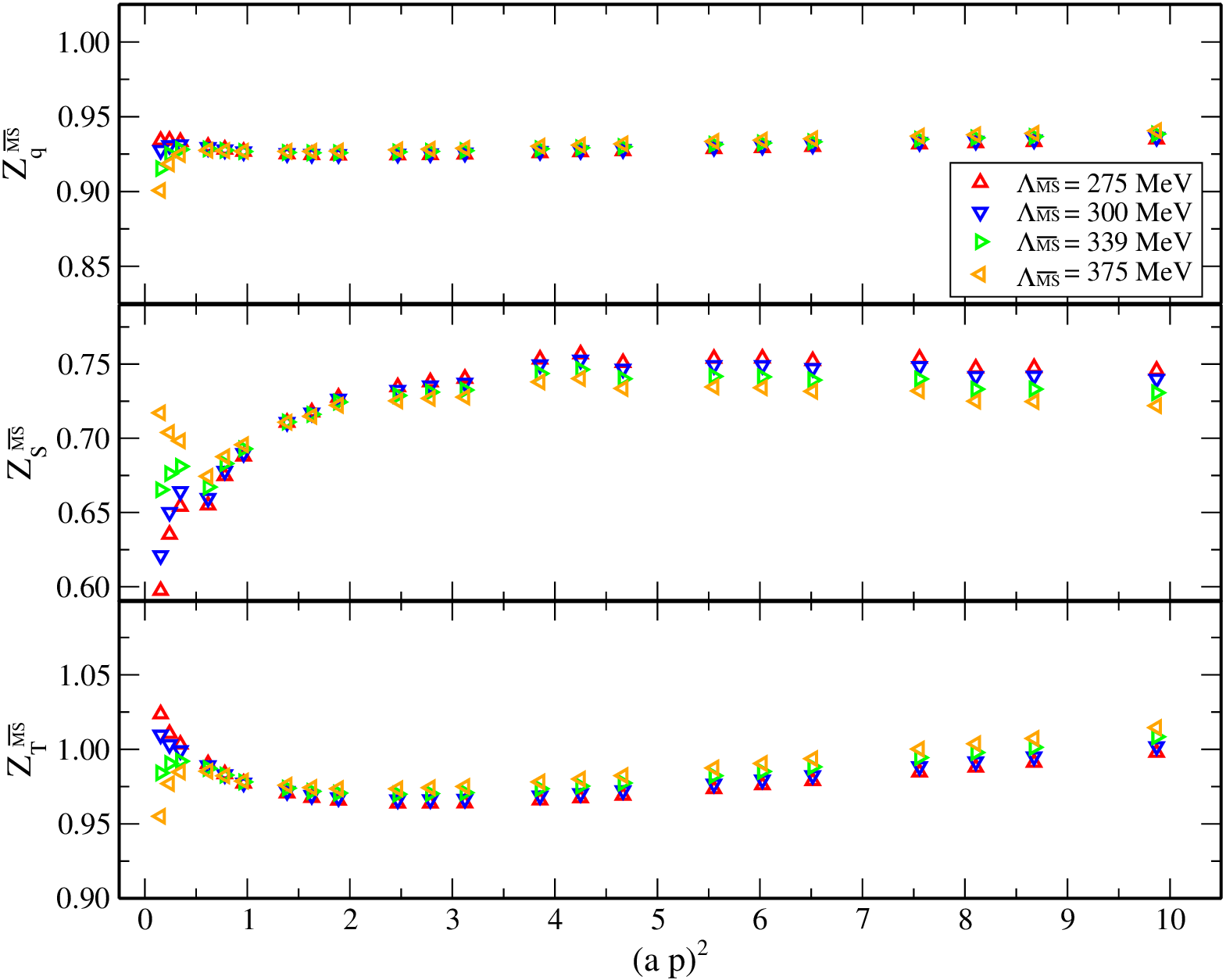}}
\caption{$Z_q,\,Z_S,\,Z_T$ in the ${\overline{\rm MS}}$ scheme using
different values of $\Lambda_{\overline{\rm MS}}$.}
\label{Zq_Zs_Zt_LQCD}
\end{figure}

Another systematic error could result from the ambiguity in the value of
$\Lambda_{\overline{\rm MS}}$ needed in Eq.~(\ref{gS}). Having this in
mind, it is interesting to see how the conversion factors are affected
by any variation of the value of $\Lambda_{\overline{\rm MS}}$. This
is demonstrated in Fig.~\ref{Zq_Zs_Zt_LQCD}. We observed that the
conversion factors of $Z_q$ and $Z_T$ are not affected by variations
of $\Lambda_{\overline{\rm MS}}$. On the other hand, $Z_S$ shows some
sensitivity on $\Lambda_{\overline{\rm MS}}$ in the range form 275 to
375 MeV. Nevertheless, this dependence on $\Lambda_{\overline{\rm MS}}$ 
is smaller than the one due to the order of $g^S(\mu)$ used in the
conversion factors.
\FloatBarrier

\subsection{Chiral extrapolation of $Z_P$ and $Z_P/Z_S$}
\label{subsec43}

\begin{center}
\begin{figure}[!h]
{\includegraphics[scale=0.4]{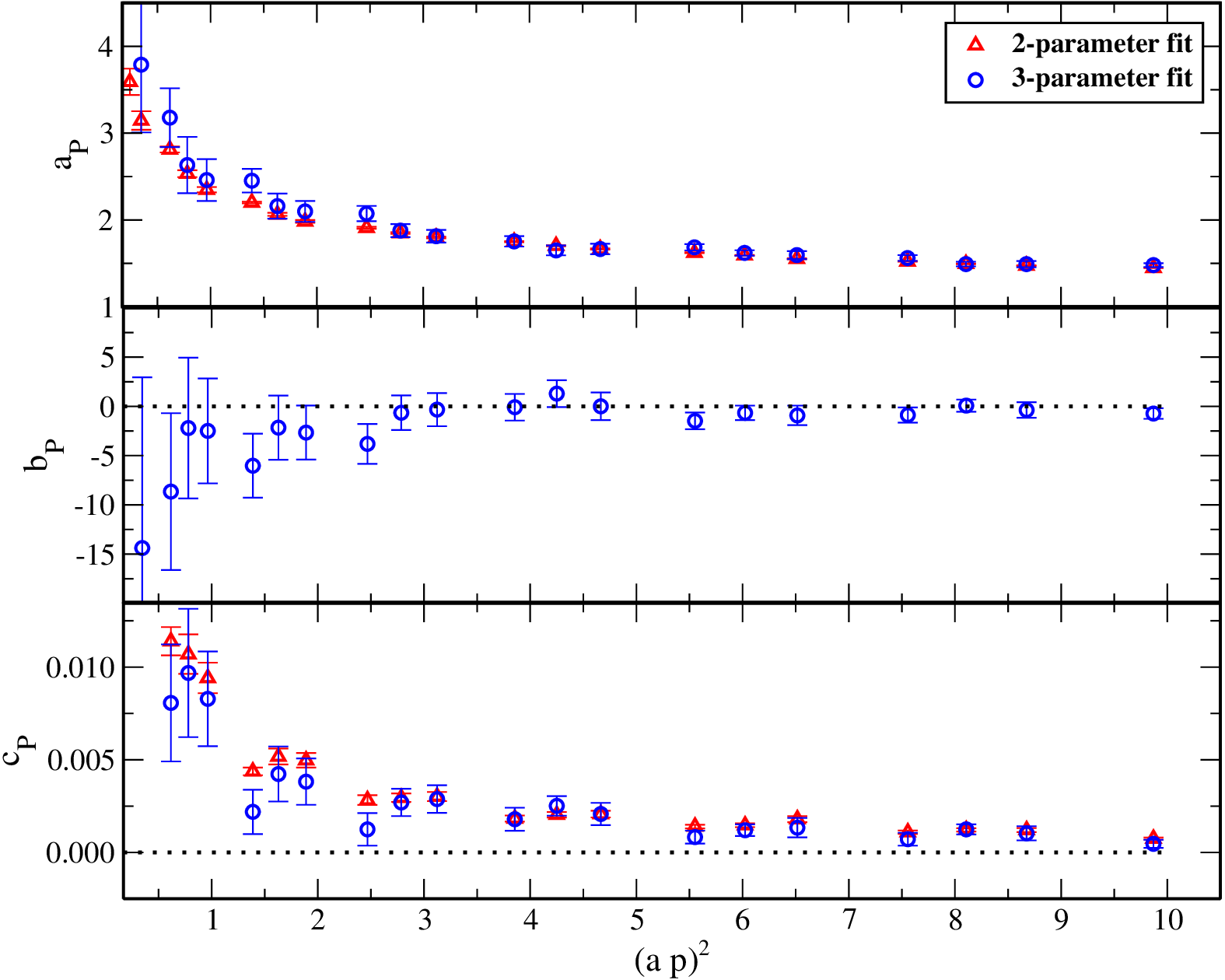}}
\caption{Red triangles: The parameters $a_P,\,c_P$ extracted from the
  2-parameter fit. Blue circles: The parameters $a_P,\,b_P,\,c_P$
  extracted from the 3-parameter fit.}
\label{2_3fit_params}
\end{figure}
\end{center}

To obtain $Z_P$ in the chiral limit, we must subtract the pion pole from the pseudoscalar 
vertex function $\Lambda_P(p,m_\pi)$. To do so we consider as  ansatz a two-parameter fit function 
for each momentum $p$
\begin{equation}
f^{(2)}(p,\,m_\pi) = a_{P}(p) + \frac{c_{P}(p)}{m_\pi^2} 
\label{PionPole2}
\end{equation}
or a three-parameter fit function
\begin{equation}
f^{(3)}(p,\,m_\pi) = a_{P}(p) + b_{P}(p)\,m_\pi^2 + \frac{c_{P}(p)}{m_\pi^2}\,,
\label{PionPole}
\end{equation}
and fit Eq. (\ref{PionPole2}) or Eq. (\ref{PionPole}) to the ratio
\be
R(p,\,m_\pi) =
\frac{\Lambda_P(p,\,m_\pi)}{Z_q(p,\,m_\pi)\,
C_{P,\rm RGI}^{\mbox{\scriptsize RI$^{\prime}$-MOM},\,{\overline{\rm MS}}} (p,2\, \rm{GeV})}\,.
\label{fit_ratio}
\ee
The coefficient $a_P(p)$ is the number we are looking for, 
the continuum limit of which corresponds to 
$(Z_P^{\overline{\rm MS}})^{-1}$ . 
In Fig.~\ref{2_3fit_params} we show the result of both local fits. The coefficient $b_P(p)$ turns 
out to be rather small. In fact, it is compatible with zero.
The two-parameter fit of Eq. (\ref{PionPole2}) leads to significantly smaller errors on $a_P(p)$. 
In the following we shall employ the local two-parameter fit to subtract the pion pole and extrapolate $Z_P$ to the chiral limit. 
The results will be shown in Sec.~\ref{sec5}. The stability of the fit 
is discussed in 
Appendix~\ref{appC} using alternatives  for the pion pole subtraction, 
along with an assessment of systematic errors.

In some applications a precise value of $Z_P/Z_S$ is needed, which suggests a direct fit of the ratio. 
In this ratio the factors $Z_q$ and $C_{\cal O}^{\mbox{\scriptsize RI$^{\prime}$-MOM},\,{\overline{\rm MS}}}$ drop out. 
To subtract the pion pole from $\Lambda_P(p,m_\pi)$, we proceed as before and 
fit Eq. (\ref{PionPole2}) and Eq. (\ref{PionPole}), respectively, with 
coefficients $a_{PS}(p)$, $b_{PS}(p)$ and $c_{PS}(p)$ to
\begin{equation}
R(p,m_\pi) = \frac{\Lambda_P(p,m_\pi)}{\Lambda_S(p,m_\pi)} \,.
\label{fit_ratio2}
\end{equation}
In the chiral limit $Z_S/Z_P=a_{PS}(p)$. 
Again, the parameter $b_{PS}(p)$ has little 
effect on the result. 
As before, we shall adopt the local two-parameter fit and show results
of the continuum extrapolation in Sec.~\ref{sec5}. 


\FloatBarrier

\section{Continuum extrapolations}
\label{sec5}

We now come to the main topic of this paper, the subtraction of lattice artifacts. Clearly, the renormalization factors $Z_V$, $Z_A$ and $Z_S^{\overline{MS}}$, $Z_P^{\overline{MS}}$ and $Z_T^{\overline{MS}}$ show some residual dependence on $(a\,p)^2$, which we will address now. The final aim is to extrapolate the data to $(a\,p)^2 = 0$. 

\subsection{Unsubtracted data}
\label{sec5a}

\begin{figure}[!h]
{\includegraphics[scale=0.5]{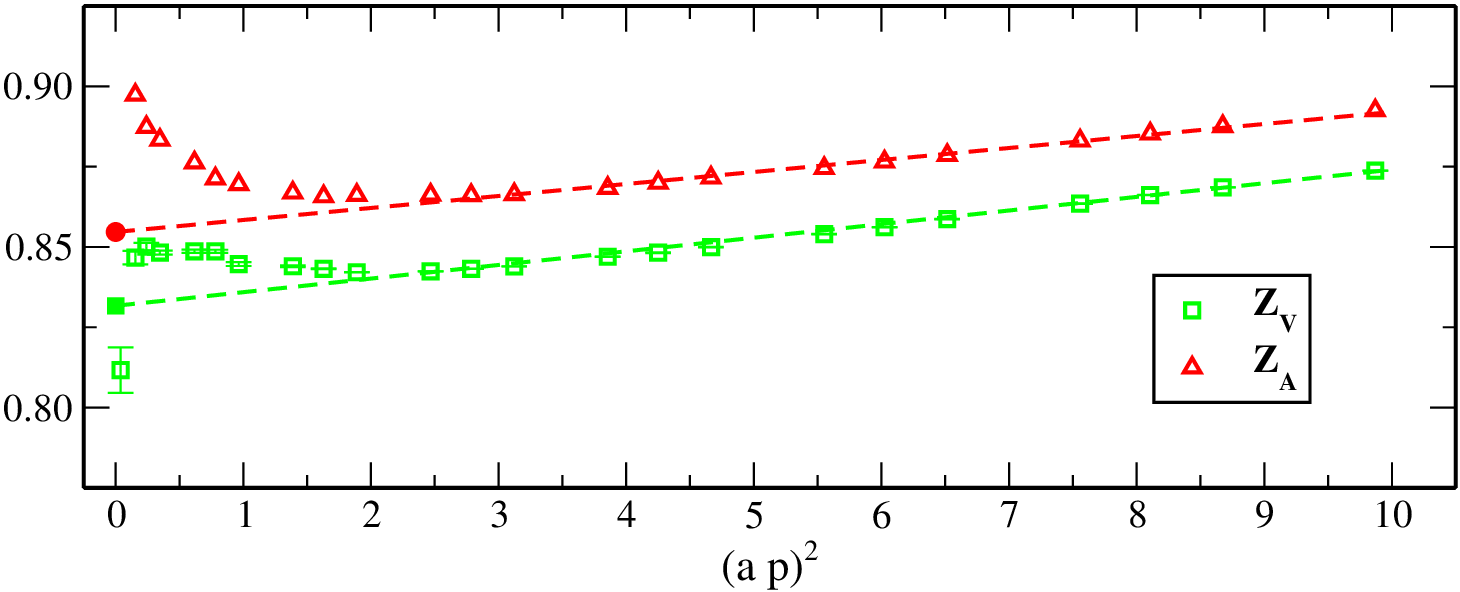}}
\caption{The dependence of $Z_V$ and $Z_A$ on the renormalization
  scale, for the chirally extrapolated data. Dashed lines represent
  the continuum extrapolation and filled points the extrapolated value.}
\label{ZvZa}
\end{figure}

\begin{figure}[!h]
{\includegraphics[scale=0.5]{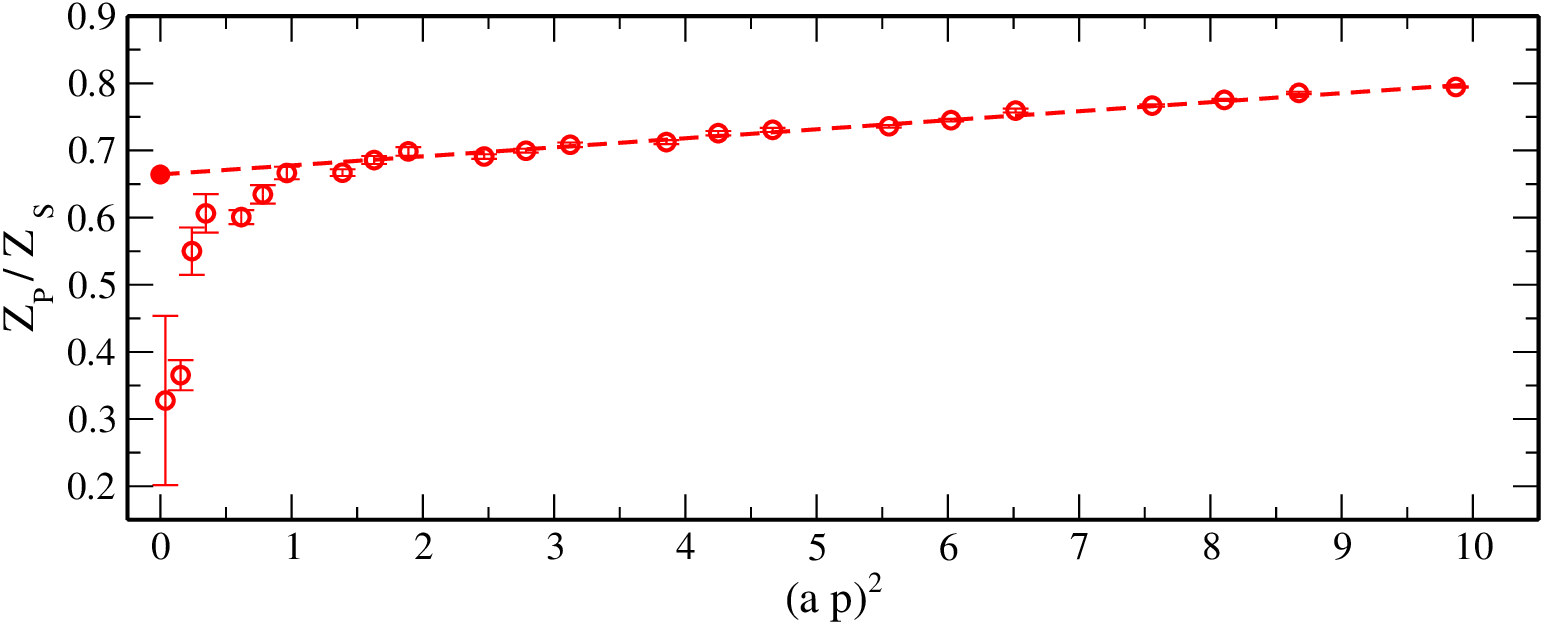}}
\caption{Similar to Fig.~\ref{ZvZa} for $Z_P$.}
\label{ZpoZs}
\end{figure}

\begin{figure}[!h]
{\includegraphics[scale=0.5]{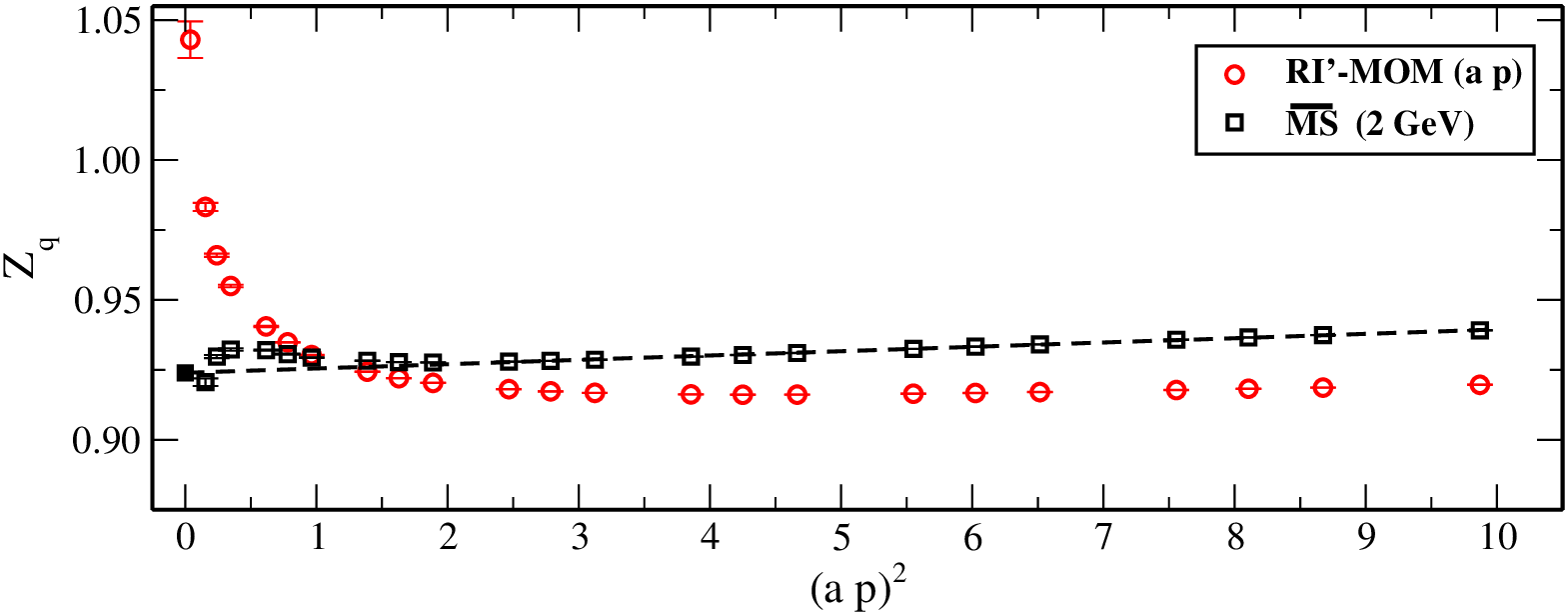}}
\caption{The dependence of $Z_q^{\mbox{\scriptsize RI$^{\prime}$-MOM}}$ and $Z_q^{\overline{\rm MS}}$ on
  the momentum scale, for the chirally extrapolated non-perturbative
  data. Dashed lines represent the continuum extrapolation and filled
  points the extrapolated value.}
\label{Zq}
\end{figure}

\begin{figure}[!h]
{\includegraphics[scale=0.5]{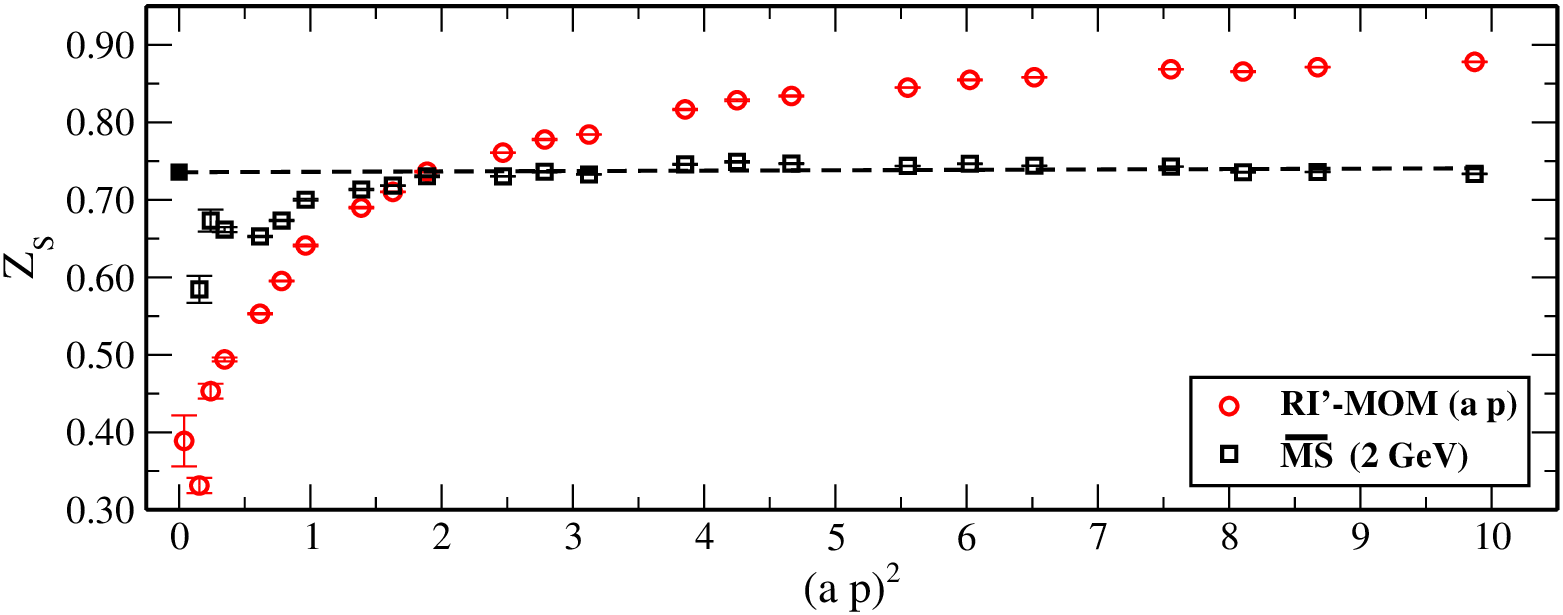}}
\caption{Similar to Fig.~\ref{Zq} for $Z_S$.}
\label{Zs}
\end{figure}

\begin{figure}[!h]
{\includegraphics[scale=0.5]{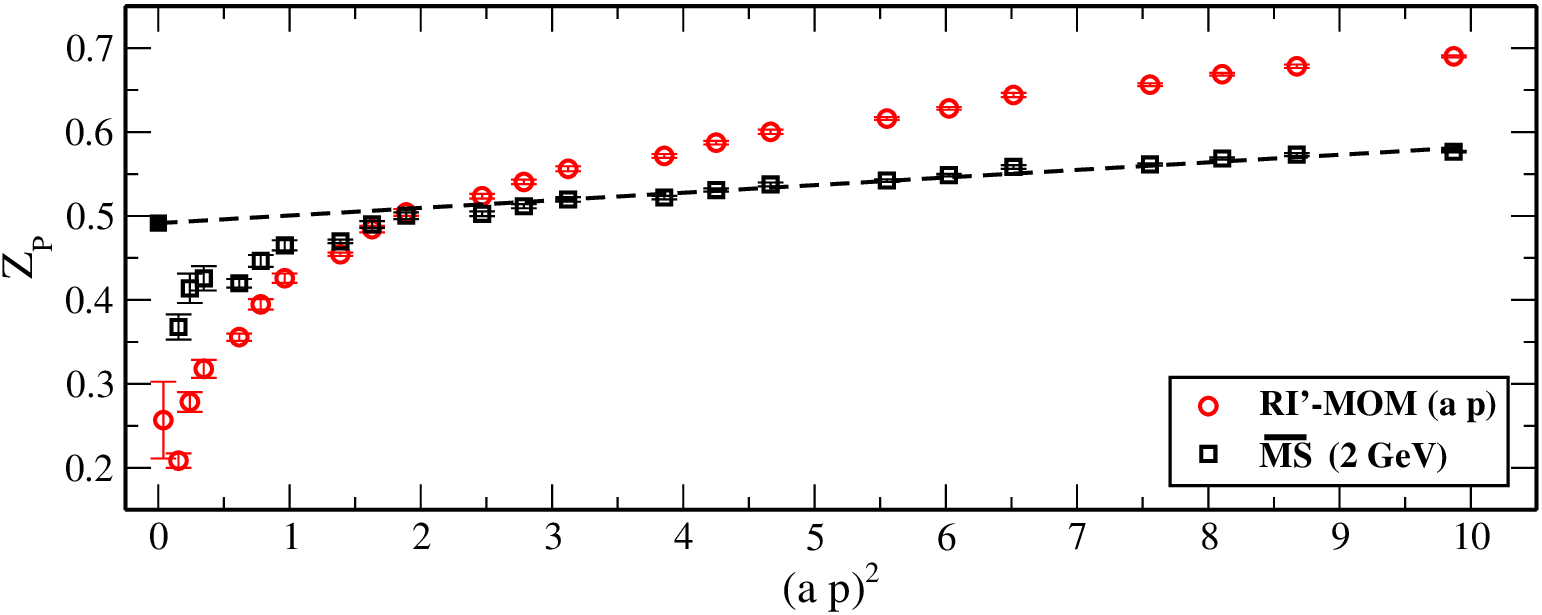}}
\caption{Similar to Fig.~\ref{Zq} for $Z_P$.}
\label{Zp}
\end{figure}

\begin{figure}[!h]
{\includegraphics[scale=0.5]{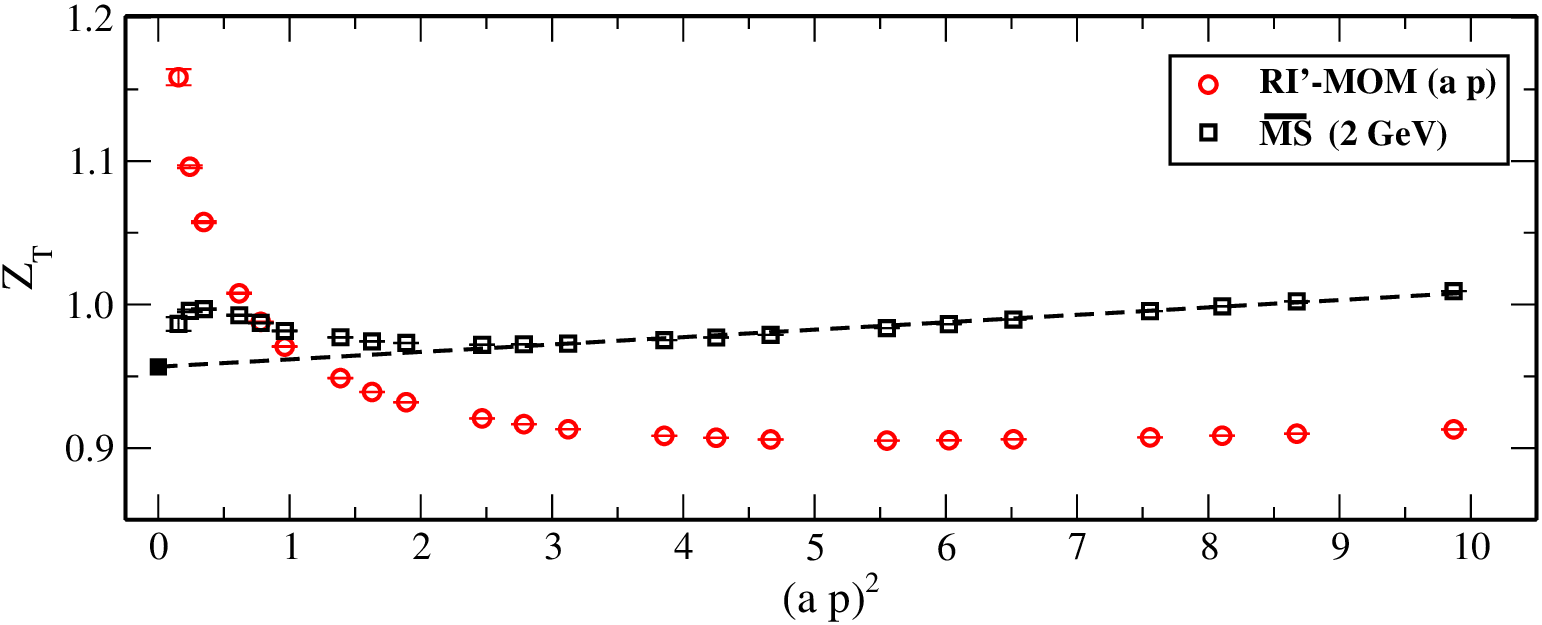}}
\caption{Similar to Fig.~\ref{Zq} for $Z_T$.}
\label{Zt}
\end{figure}

Let us first look at the raw data, extrapolated to the chiral limit and converted to the $\overline{MS}$ scheme 
in Sec.~\ref{sec4}.
%
%
In Figs.~\ref{ZvZa} - \ref{ZpoZs} we plot $Z_V$, $Z_A$ and $Z_P/Z_S$ as a function of $(a\,p)^2$. 
Similarly, in Figs.~\ref{Zq}, \ref{Zs}, \ref{Zp} and \ref{Zt} we plot $Z_q$, $Z_S$, $Z_P$ 
and $Z_T$ in the RI$'$-MOM and $\overline{\rm MS}$ scheme, respectively, at $\mu=2\,{\rm GeV}$. 
We find $Z_V$, $Z_A$ and $Z_P/Z_S$ and $Z_S^{\overline{MS}}$, $Z_P^{\overline{MS}}$ and 
$Z_T^{\overline{MS}}$ to lie approximately on a linear curve for $(a\,p)^2 \gtrsim 2$, 
which allows a fit to a straight line. The dashed lines show a fit to the interval $(a\,p)^2 \in [2,10]$. 

While statistical errors are small, there are some systematic errors, which should 
be carefully examined. One source of error is the accuracy of the
conversion factors. Another source arises from the choice of fit interval. We 
have also done fits to the intervals $[1,10]$, $[3.7,10]$ and $[2,6]$. 
The difference in results will give us an estimate of the systematic error.

\FloatBarrier 

\FloatBarrier 

\FloatBarrier
\FloatBarrier
\FloatBarrier
\FloatBarrier

In Table~\ref{tab_all_final} we present the $Z$-factors, after chiral and continuum extrapolation, 
from a fit to the interval $(a\,p)^2 \in [2,10]$. The number in the first bracket indicates the 
purely statistical error. The number in the second bracket states the systematic error, which is 
taken from the difference of the fit to $(a\,p)^2 \in [2,10]$ and $[2,6]$. The third number, 
wherever it applies, reflects the difference of two- and three-loop conversion factors.

\begin{table}[!h]
\begin{center}
\begin{tabular}{|c|c|}
\hline
$Z_q^{\overline{\rm MS}}$(2 GeV)   &  $\,\,\,$0.9239(001)(003)(028)$\,\,\,$              \\
$Z_S^{\overline{\rm MS}}$(2 GeV)   &  $\,\,\,$0.7356(053)(155)(277)$\,\,\,$              \\
$Z_P^{\overline{\rm MS}}$(2 GeV)   &  $\,\,\,$0.4915(025)(125)(184)$\,\,\,$              \\
$Z_P/Z_S$                         &  $\,\,\,$0.6643(032)(023)\phantom{(000)}$\,\,\,$ \\
$Z_V$                             &  $\,\,\,$0.8317(005)(021)\phantom{(000)}$\,\,\,$ \\
$Z_A$                             &  $\,\,\,$0.8547(0008)(036)\phantom{(000)}$\,\,\,$ \\
$Z_T^{\overline{\rm MS}}$(2 GeV)   &  $\,\,\,$0.9566(013)(054)(101)$\,\,\,$               \\
\hline
\end{tabular}
\end{center}
\caption{Continuum extrapolated values on $Z_q^{\overline{\rm MS}}$,
  $Z_S^{\overline{\rm MS}}$, $Z_P^{\overline{\rm MS}}$,
  $Z_T^{\overline{\rm MS}}$, $Z_V$, $Z_A$ and $Z_P/Z_S$. The number in
  the first (second) bracket is the statistical (systematic) error,
  and where applicable, the one in the third bracket comes from
  the difference in using the two- and three-loop results for the
  conversion factor via RGI.}
\label{tab_all_final}
\end{table}
\FloatBarrier

\subsection{Subtraction of one-loop perturbative lattice artifacts}
\label{sec5b}

\begin{figure}[!h]
{\includegraphics[scale=0.5]{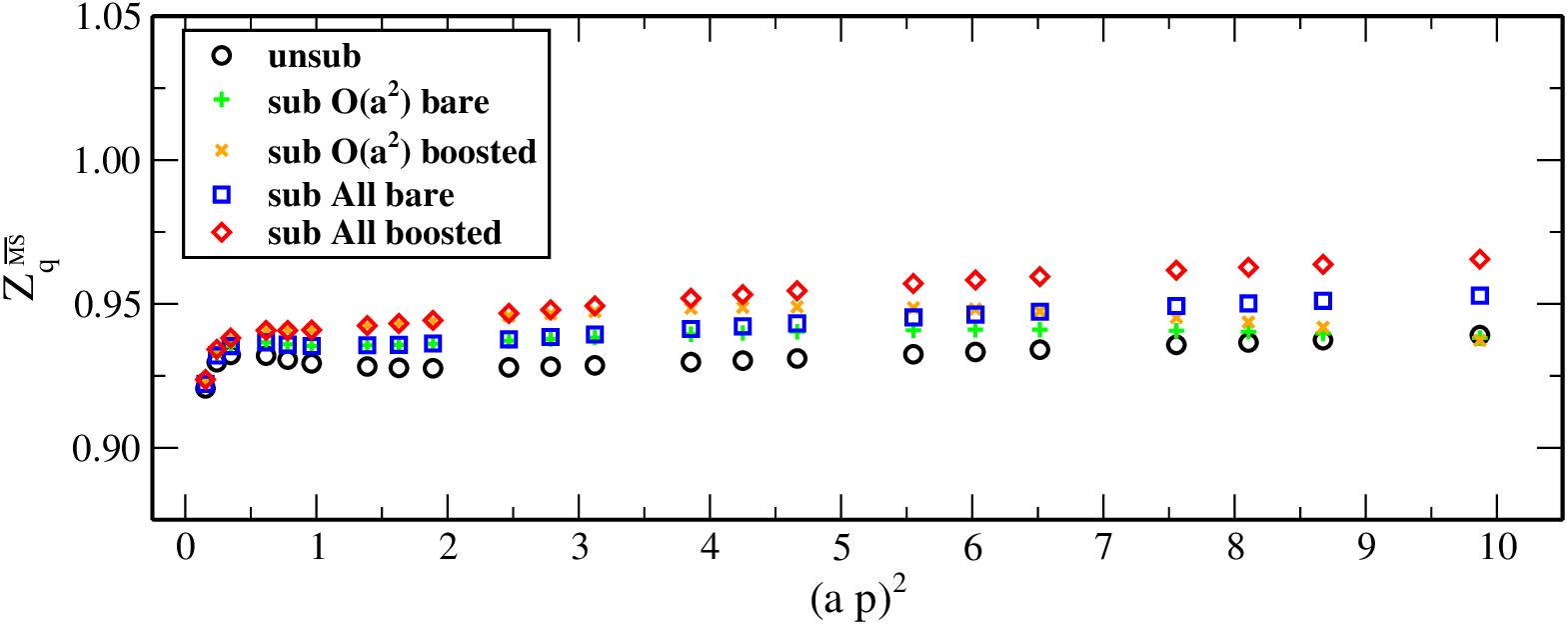}}
\vskip -0.3cm
\caption{Chirally extrapolated values for $Z_q^{\overline{\rm MS}}$
  before the perturbative subtraction (black circles) and after subtraction of:
  a. ${\cal O}(g^2\,a^2)$ terms using bare coupling, $g$ (green plus points),
  b. ${\cal O}(g_b^2\,a^2)$ terms using boosted coupling, $g_b$ (orange crosses),
  c. complete subtraction using $g$ (blue squares), and
  d. complete subtraction using $g_b$ (red diamonds). }
\label{Zqsub}
\end{figure}

\begin{figure}[!h]
{\includegraphics[scale=0.5]{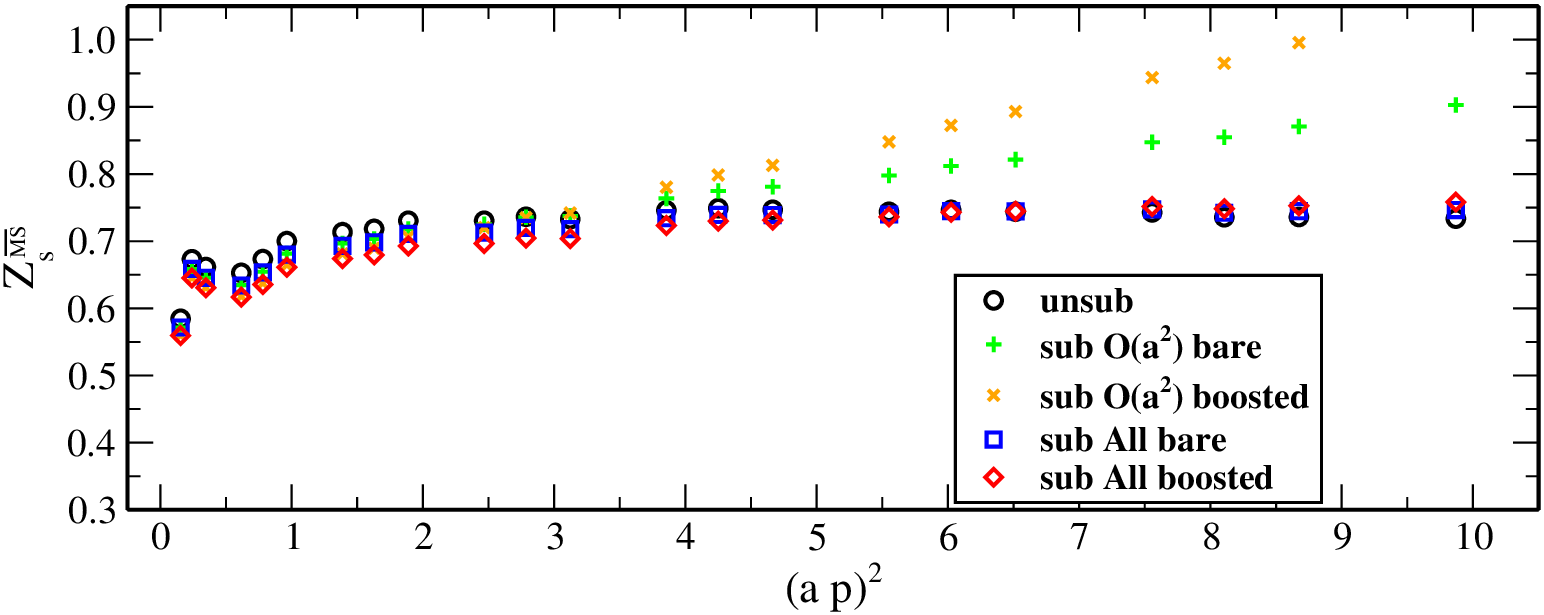}}
\vskip -0.3cm
\caption{Similar to Fig.~\ref{Zqsub} for $Z_S$.}
\label{Zssub}
\end{figure}

\begin{figure}[!h]
{\includegraphics[scale=0.5]{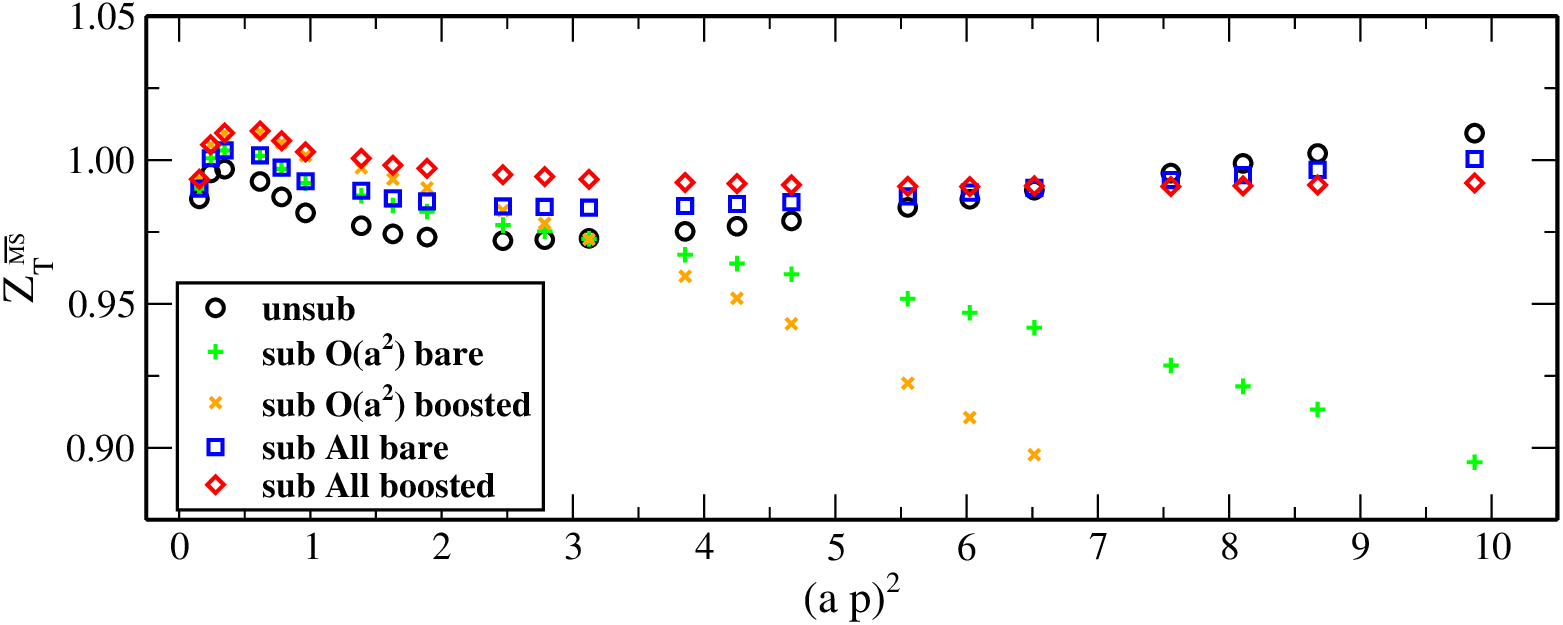}}
\vskip -0.3cm
\caption{Similar to Fig.~\ref{Zqsub} for $Z_T$.}
\label{Ztsub}
\end{figure}

We now turn to the subtraction of lattice artifacts.
There are two kinds of subtractions we employ (see Sec.~\ref{sec3}), the subtraction of one-loop ${\cal O}(a^2)$ 
corrections and the complete
subtraction of one-loop lattice artifacts. For each case we use both the
bare, $g$, and boosted coupling, 
\be
g_b^2 = \frac{g^2}{P(g)}\,,
\ee
where $P(g)$ is the plaquette at $\beta=5.50$ and for the                                                   
ensembles used in this work we find: $P(\beta=5.5) \sim 0.52$.
For the improvement coefficient $c_{\rm sw}$ we take the tree-level value, $c_{\rm sw}=1$. 
The effect of the subtraction is demonstrated as examples in Figs.~\ref{Zqsub} -
\ref{Ztsub} for the chirally extrapolated data. 
For $Z_q$ (Fig.~\ref{Zqsub}) we find no clear preference for any kind of subtraction. 
In case of the quark-bilinear operators (Figs.~\ref{Zssub} -\ref{Ztsub}) the ${\cal O}(g^2\,a^2)$ 
corrections are very small for $(a\,p)^2 < 4$, 
but beyond that they get out of control and show significant $O((a\,p)^4)$ effects. Complete 
subtraction of one-loop lattice artifacts, on the other hand, has a small, albeit appreciable, 
effect on the $Z$-factors. In case of $Z_V,\,Z_A$ (not shown here) 
and $Z_T$ 
the data is brought on a perfectly straight line for $(a\,p)^2 \in [2,10]$, using the 
boosted coupling $g_b^2$.

\vspace{0.3cm}
\FloatBarrier
\FloatBarrier
%
%
%
%
%
\FloatBarrier

\section{Final results and discussion}
\label{sec6}

\begin{figure}[!h]
{\includegraphics[scale=0.485]{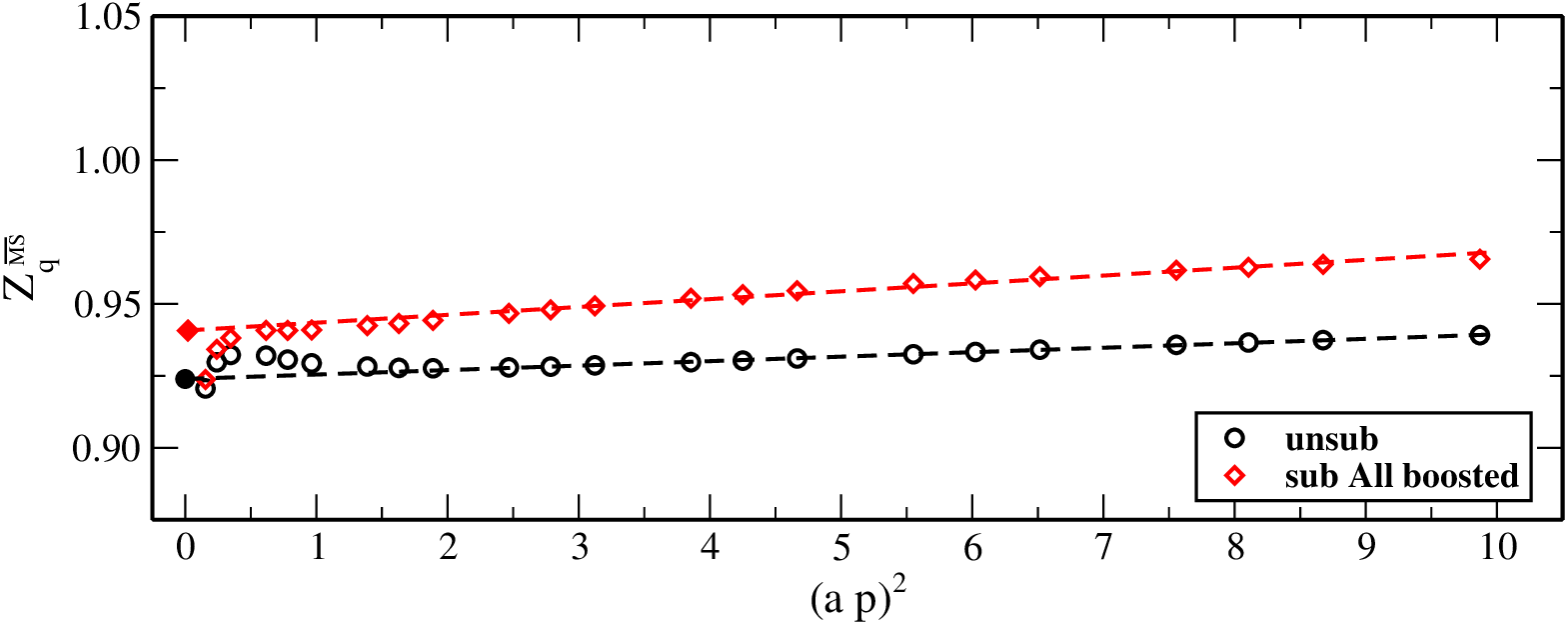}}
\vskip -0.5cm
\caption{Chirally extrapolated values for $Z_q^{\overline{\rm MS}}$
  prior the perturbative subtraction (black circles) and after the
  complete subtraction of one-loop lattice artifacts using $g_b$ (red diamonds). }
\label{Zqsub2}
\end{figure}

\begin{figure}[!h]
{\includegraphics[scale=0.485]{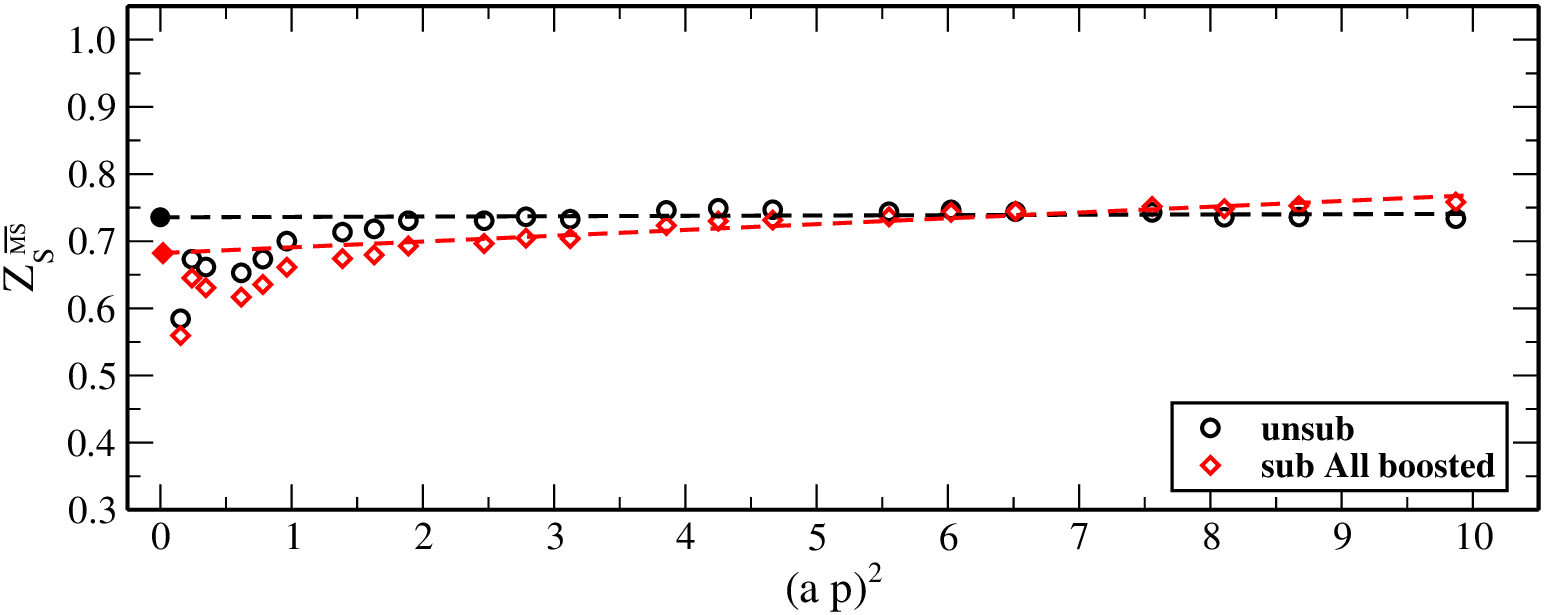}}
\vskip -0.6cm
\caption{Similar to Fig.~\ref{Zqsub2} for $Z_S$.}
\label{Zssub2}
\end{figure}

\begin{figure}[!h]
{\includegraphics[scale=0.485]{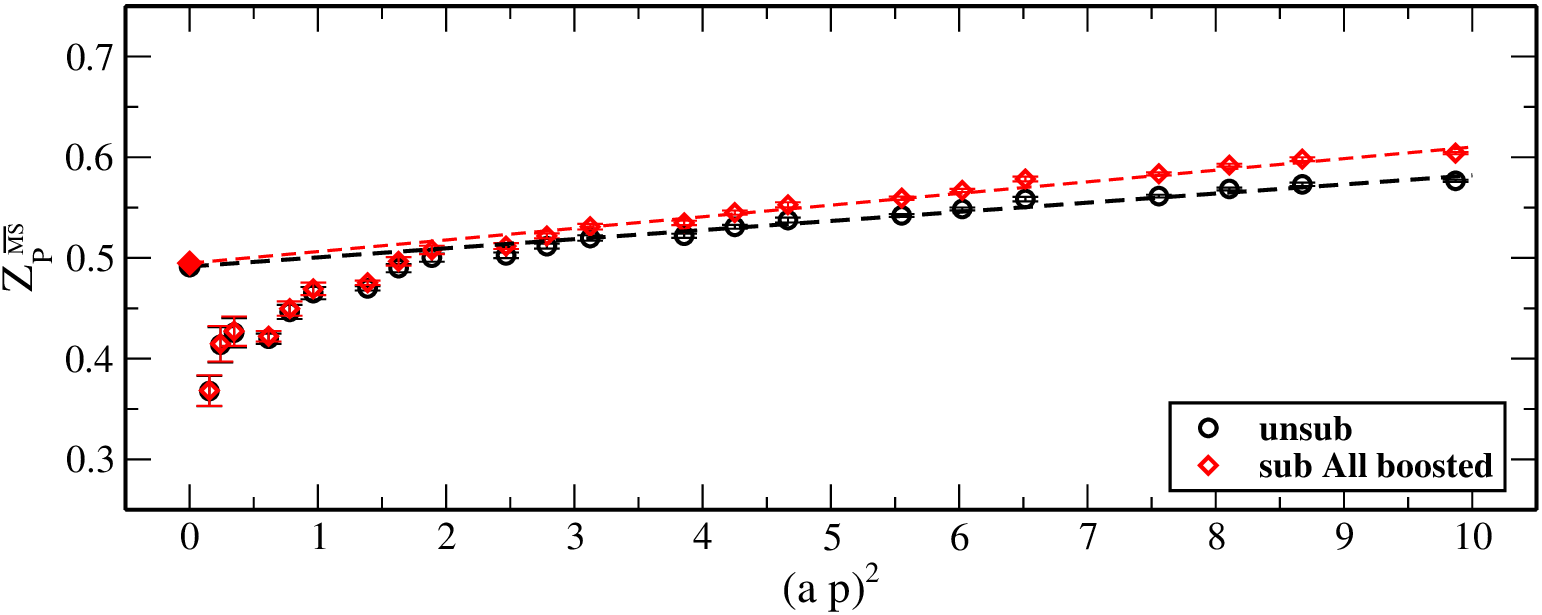}}
\vskip -0.6cm
\caption{Similar to Fig.~\ref{Zqsub2} for $Z_P$.}
\label{Zpsub2}
\end{figure}

\begin{figure}[!h]
{\includegraphics[scale=0.485]{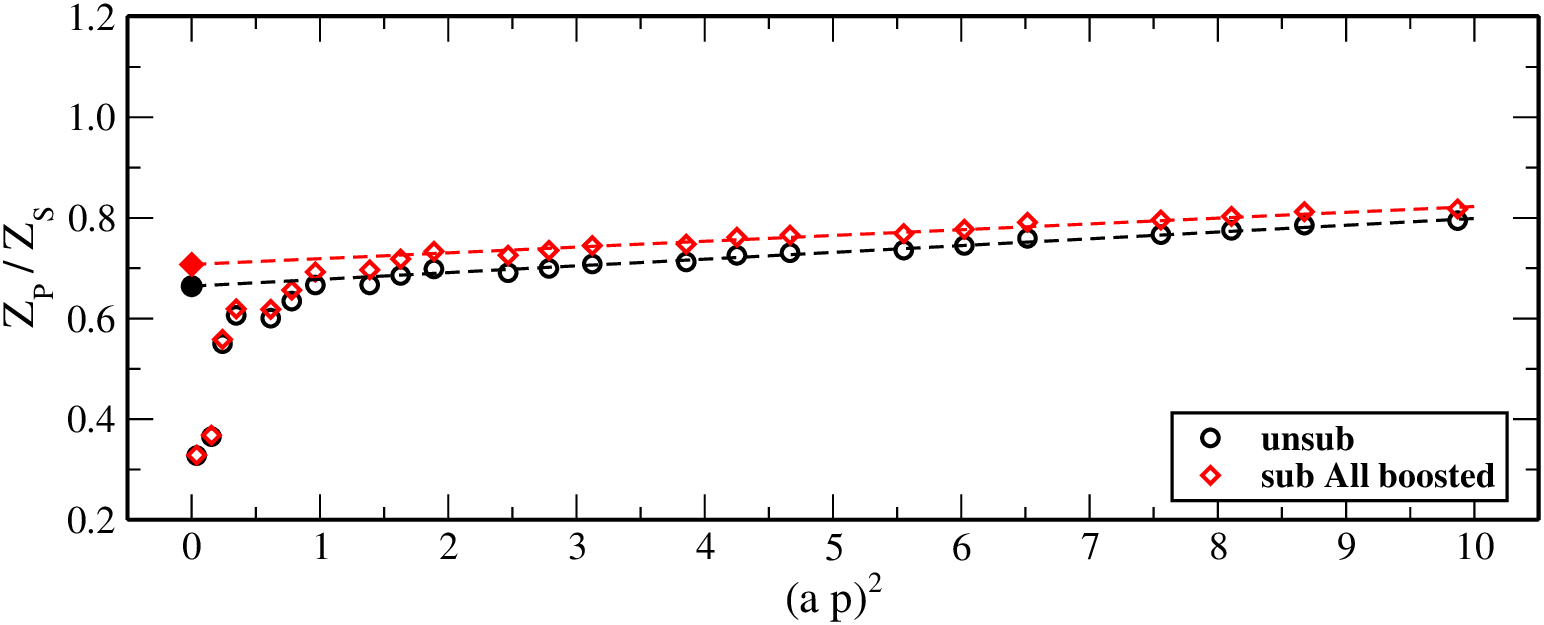}}
\vskip -0.3cm
\caption{Similar to Fig.~\ref{Zqsub2} for $Z_P/Z_S$.}
\label{ZpoZssub2}
\end{figure}

\begin{figure}[!h]
{\includegraphics[scale=0.485]{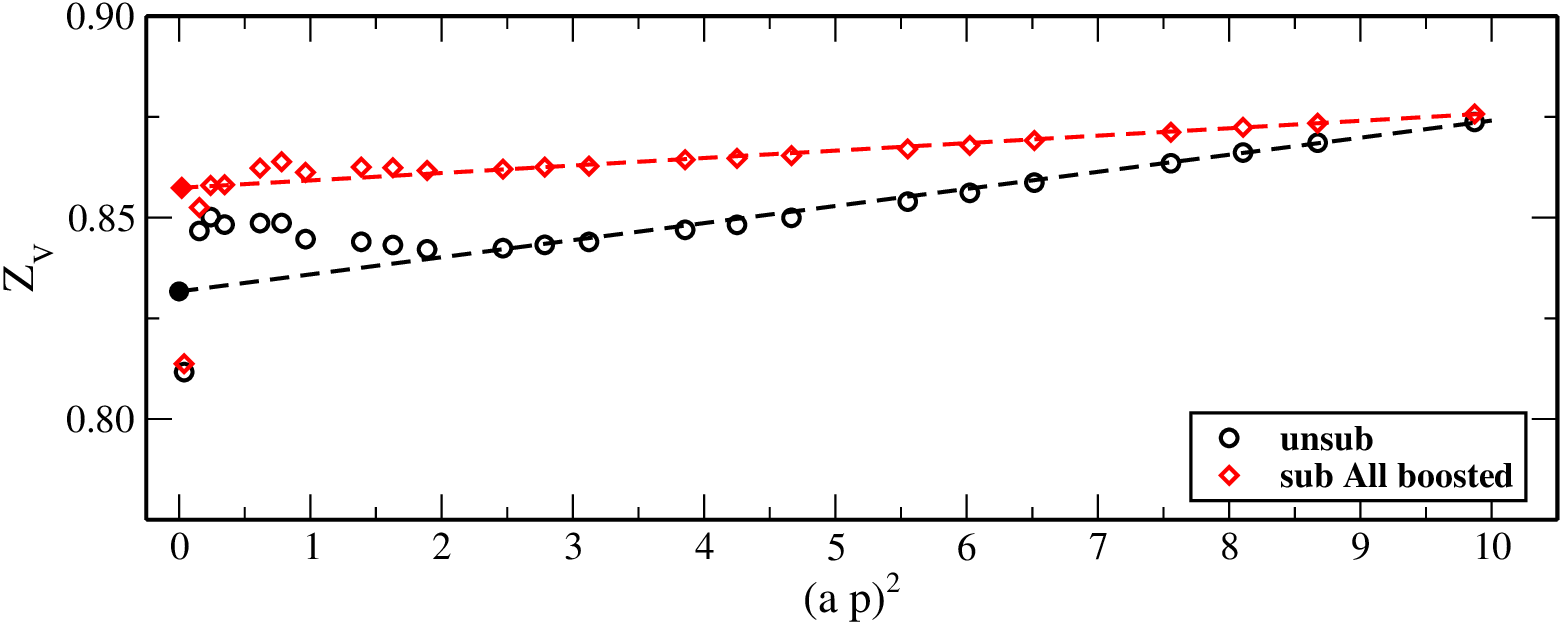}}
\vskip -0.3cm
\caption{Similar to Fig.~\ref{Zqsub2} for $Z_V$.}
\label{Zvsub2}
\end{figure}

\begin{figure}[!h]
{\includegraphics[scale=0.485]{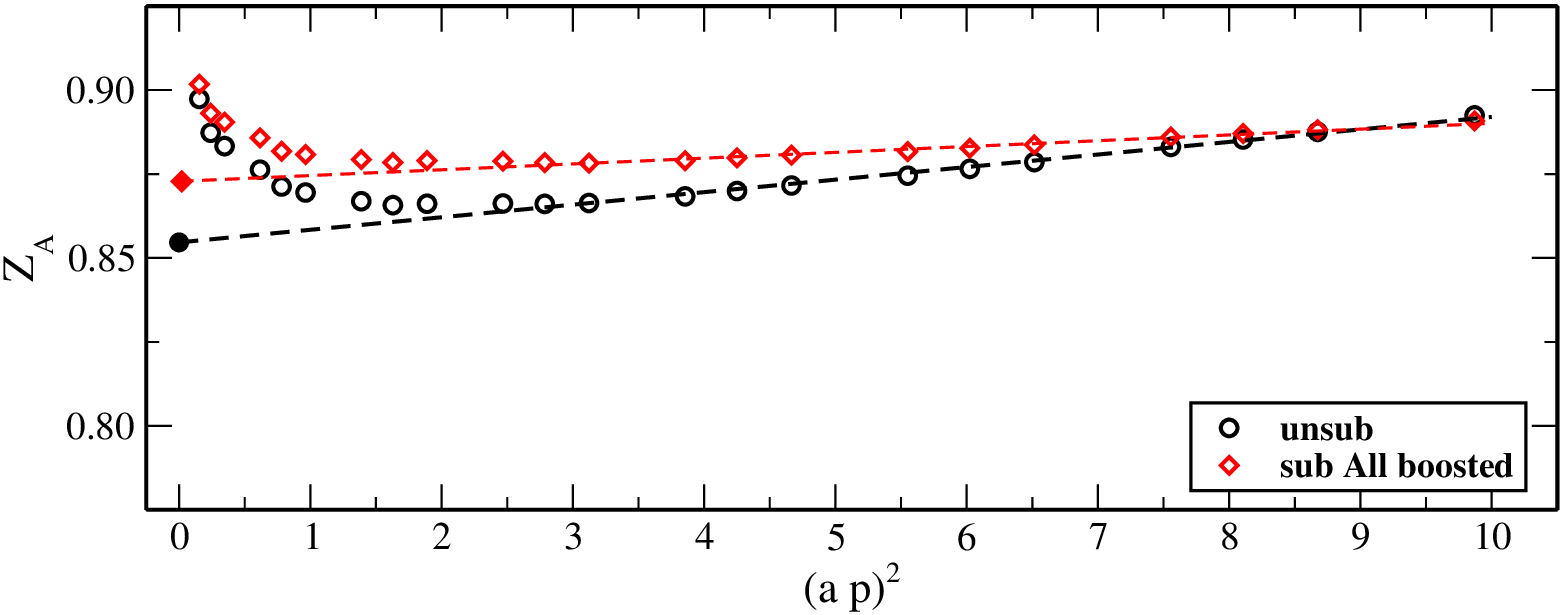}}
\vskip -0.3cm
\caption{Similar to Fig.~\ref{Zqsub2} for $Z_A$.}
\label{Zasub2}
\end{figure}

\begin{figure}[!h]
{\includegraphics[scale=0.485]{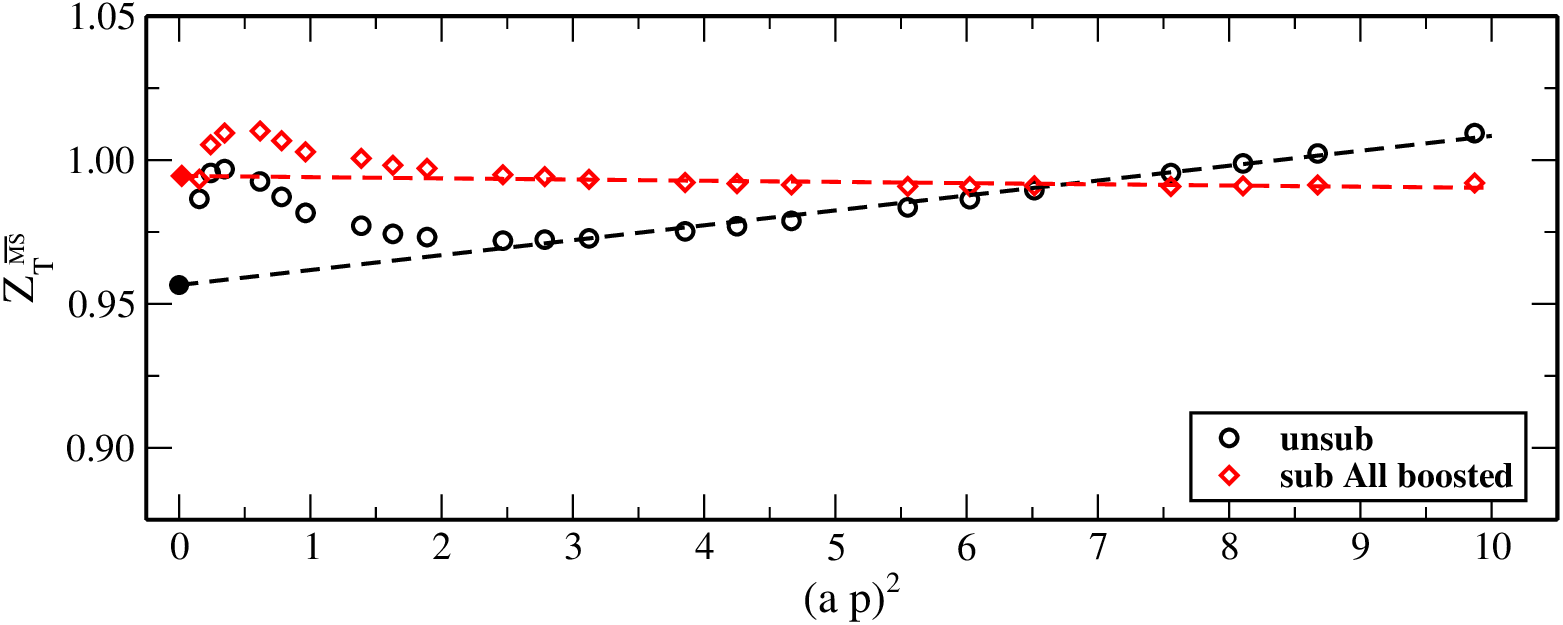}}
\vskip -0.3cm
\caption{Similar to Fig.~\ref{Zqsub2} for $Z_T$.}
\label{Ztsub2}
\end{figure}

We take the lattice data of Sec.~\ref{sec4}, improved by complete subtraction of one-loop lattice 
artifacts with boosted coupling $g_b$, as our final result. In Figs.~\ref{Zqsub2} - \ref{Ztsub2} 
we show the $Z$-factors before and after the subtraction, together with a linear extrapolation 
to the continuum.
\FloatBarrier
\vskip -0.2cm
\FloatBarrier
\vskip -0.2cm
\FloatBarrier

\FloatBarrier

\FloatBarrier

\FloatBarrier

\FloatBarrier

In Table~\ref{tab_all_sub} we give our final numbers, corresponding to the solid diamonds in Figs.~\ref{Zqsub2} - \ref{Ztsub2}. The corrected numbers differ by up to $8\%$ from the unsubtracted results in Table~\ref{tab_all_final}.

\begin{table}[!h]
\begin{center}
\begin{tabular}{|c|c|}
\hline
$Z_q^{\overline{\rm MS}}$      & $\,\,\,$ 0.9408(008)(024)$\,\,\,$  \\
$Z_S^{\overline{\rm MS}}$      & $\,\,\,$ 0.6822(061)(176)$\,\,\,$  \\
$Z_P^{\overline{\rm MS}}$      & $\,\,\,$ 0.4948(026)(128)$\,\,\,$  \\
$Z_P/Z_S$                & $\,\,\,$ 0.7075(036)(068)$\,\,\,$  \\
$Z_V$                      & $\,\,\,$ 0.8574(002)(007)$\,\,\,$  \\
$Z_A$                      & $\,\,\,$ 0.8728(006)(027)$\,\,\,$  \\
$Z_T^{\overline{\rm MS}}$      & $\,\,\,$ 0.9945(010)(035)$\,\,\,$  \\
\hline
\end{tabular}
\end{center}
\caption{Continuum results for $Z_q$, $Z_S$, $Z_P$, $Z_P/Z_S$, $Z_V$, $Z_A$
and $Z_T$ in the $\overline{\rm MS}$ scheme at $\mu=2\,\mbox{GeV}$, where it applies. The numbers refer to the fit interval $(a\,p)^2 \in [2,10]$. The first number in brackets is the statistical error, the second number the systematic error due to the fit range.}
\label{tab_all_sub}
\end{table}
\FloatBarrier

The conversion factors to the RGI scheme and the RI$'$-MOM scheme at $\mu=1/a$ are given by
\begin{equation}
\begin{array}{lclclcl}
Z_q^{\rm RGI}    &=& 0.9461 \, Z_q^{\overline{\rm MS}}\,, & \phantom{0000} &
Z_q^{\mbox{\scriptsize RI$^{\prime}$-MOM}}    &=& 1.0004 \, Z_q^{\overline{\rm MS}}\,, \\
Z_S^{\rm RGI}    &=& 0.7503 \, Z_S^{\overline{\rm MS}}\,, & \phantom{0000} &
Z_S^{\mbox{\scriptsize RI$^{\prime}$-MOM}}    &=& 0.9211 \, Z_S^{\overline{\rm MS}}\,, \\
Z_P^{\rm RGI}    &=& 0.7503 \, Z_P^{\overline{\rm MS}}\,, & \phantom{0000} &
Z_P^{\mbox{\scriptsize RI$^{\prime}$-MOM}}    &=& 0.9211 \, Z_P^{\overline{\rm MS}}\,, \\
Z_T^{\rm RGI}    &=& 1.0604 \, Z_T^{\overline{\rm MS}}\,, & \phantom{0000} &
Z_T^{\mbox{\scriptsize RI$^{\prime}$-MOM}}    &=& 0.9870 \, Z_T^{\overline{\rm MS}}\,.
\end{array}
\end{equation}


To recapitulate, in a previous work~\cite{Constantinou:2013ada} we
have computed renormalization factors of quark-bilinear operators for
$N_f=2$ flavors of dynamical clover fermions. We have refined the
original procedure~\cite{Martinelli:1994ty} in several aspects,
including the use of momentum sources and the perturbative subtraction
of lattice artifacts. In this work we extended the calculation to
$N_f=3$ flavors of SLiNC fermions~\cite{Bietenholz:2011qq}. Using
twisted boundary conditions, the lattice momenta were chosen to lie
strictly on the diagonal, $\displaystyle p_\mu= \sqrt{p^2/4},\,
\forall \mu$. 

Complete subtraction of one-loop lattice artifacts has 
brought $Z_V$, $Z_A$, $Z_S^{\overline{MS}}$, $Z_P^{\overline{MS}}$ and
$Z_T^{\overline{MS}}$ onto a straight line for $(a\,p)^2 \in [2,10]$,
\begin{equation}
Z(p) = Z(0) + Z^\prime(0)\,(a\,p)^2\,,
\end{equation}
with the continuum value being given by $Z(0)$. The origin of the
remaining $(a\,p)^2$-dependence is not completely clear to us. The
corrections are found to be substantial. In case of $Z_S$ they amount
to $8\%$. The renormalization factor of the local vector current,
$Z_V$, can be determined independently from the Dirac form factor of
the proton at zero momentum transfer by demanding that
$F_1^p(0)=1$. In~\cite{Shanahan:2014cga} we found $Z_V=0.857(1)$ at a
pion mass of $m_\pi=220\,\mbox{MeV}$, which is in perfect agreement
with our final number in Table~\ref{tab_all_sub}. This gives support
for our procedure of subtracting lattice artifacts.

It helped that the unsubtracted data were approximately linear in $(a\,p)^2$ already 
for $(a\,p)^2 \gtrsim 2$, in contrast to the case of $N_f=2$ flavors of clover fermions and 
the plaquette action~\cite{Constantinou:2013ada}.
We attribute that to the Symanzik and stout-link improved action
employed here, which appears to suppress lattice artifacts in quark
Green's functions; see e.g., Figs.~\ref{Zq_Zt_all_Oa2} -
\ref{Zv_Za_all_Oa2} demonstrating that the stout smearing leads to
smaller lattice artifacts compared to the non-smeared case ($\omega =0$).

\section*{Acknowledgements}
\vskip -0.3cm
The numerical configuration generation was performed using the BQCD
lattice QCD program, \cite{Nakamura:2010qh}, on the IBM BlueGeneQ
using DIRAC 2 resources (EPCC, Edinburgh, UK), the BlueGene P and Q at
NIC (J\"ulich, Germany) and the Cray XC30 at HLRN (Berlin and Hannover, Germany).
The BlueGene codes were optimized using Bagel~\cite{Boyle:2009vp}.
The Chroma software library~\cite{Edwards:2004sx} was used in the data analysis.
This work has been supported in part by the EU grants 227431 (Hadron Physics2), 283286 (HadronPhysics3) and by the Australian
Research Council under grants FT100100005 and DP140103067 (JMZ).
Further support was provided by the project 
TECHNOLOGY/$\Theta$E$\Pi$I$\Sigma$/0311(BE)/16
funded by the Cyprus Research Promotion Foundation,
and by Deutsche Forschungsgemeinschaft DFG Grant under contract SCHI 422/9-1.
\appendix

\newpage
\section{Perturbative Results}
\label{appA}

In this section we summarize the perturbative results for the ${\cal
  O}(g^2\,a^2)$ contributions $\Lambda_q^{(2)}$ and $\Lambda_{\cal O}^{(2)}$
as defined in Sec.~\ref{sec3} (see Eq.~(\ref{LL})). For simplicity,
we restrict to the case of the SLiNC action, which uses tree-level
Symanzik gluons. The expressions, evaluated at $\omega=0$, correspond
to the usual clover action. Our results are given as a function of the 
momentum $p$ and general values for the action parameters 
$c_{\rm sw},\,\omega$. The gauge fixing parameter $\alpha$ is 0(1) for the
Landau (Feynman) gauge. 
\bea
\Lambda_q^{(2)} = a^2\,\ggcf\,
&\hspace{-0.15cm}\Biggl(\hspace{-0.15cm}&p^2 \Bigl(1.14717
-1.51605\,\alpha + c_{\rm sw}(-0.653431+ 0.505886 \,\omega)
-0.497834 \,c_{\rm sw}^2\nonumber \\
&&\hspace{0.45cm}  -12.0983\,\omega +28.0799  \,\omega^2
+\log (a^2\,p^2) \left(-\frac{73}{360}+\frac{3 \,\alpha}{8}+\frac{c_{\rm sw}^2}{4}+\frac{c_{\rm sw}}{4}+\,\omega \right)   \Bigr) \nonumber \\
&+&
\frac{p4}{p^2}\, \Bigl(2.10650 +0.395834\,\alpha +c_{\rm sw}(0.284537 -0.362507 \,\omega ) +0.128381 c_{\rm sw}^2\nonumber \\
&&\hspace{0.45cm} -4.08165  \,\omega -16.0889 \,\omega^2 
-\frac{157}{180} \log (a^2\,p^2)   \Bigr) \\  [2ex]
\Lambda_S^{(2)} = a^2\,\ggcf\,
&\hspace{-0.15cm}\Biggl(\hspace{-0.15cm}&p^2 \Bigl(-1.20757 +0.75755
\,\alpha+c_{\rm sw}(3.19935 -4.79168 \,\omega)
-0.69430 \,c_{\rm sw}^2\nonumber \\
&&\hspace{0.45cm} -0.64987 \,\omega +0.71144\,\omega^2
+\log (a^2\,p^2) \left(\frac{17}{360}-\frac{3\,\alpha}{8}-\frac{5\,c_{\rm sw}}{4}+\frac{c_{\rm sw}^2}{4}+\,\omega\right)  \Bigr) \nonumber \\
&+&
\frac{p4}{p^2}\, \Bigl(1.6065+0.52083\,\alpha+c_{\rm sw}(0.28454 -0.36251 \,\omega )
+0.12838\,c_{\rm sw}^2\nonumber \\
&&\hspace{0.45cm}-4.08165 \,\omega-16.0889 \,\omega^2 -\frac{157}{180} \log(a^2\,p^2)\Bigr) \Biggr) \\ [2ex]
\Lambda_P^{(2)} =    a^2\,\ggcf &\hspace{-0.15cm}\Biggl(\hspace{-0.15cm}&p^2 \Bigl(0.44076-0.67794 \,\alpha +c_{\rm sw}(0.50589 \,\omega -0.65343)
-0.22227 \,c_{\rm sw}^2 \nonumber \\
&&\hspace{0.45cm}-5.59237 \,\omega -1.17320 \,\omega^2
+ \log(a^2\,p^2) \left(\frac{17}{360}+\frac{\alpha}{8}+\frac{c_{\rm sw}}{4}+\frac{c_{\rm sw}^2}{4}+\omega \right) \Bigr) \nonumber \\
&+&
\frac{p4}{p^2}\, \Bigl(1.6065+0.52083\,\alpha +c_{\rm sw}(0.28454 -0.36251 \,\omega )
+0.12838\,c_{\rm sw}^2\nonumber \\
&&\hspace{0.45cm}-4.08165 \,\omega-16.0889 \,\omega^2 -\frac{157}{180}\log(a^2\,p^2)\Bigr) \Biggr) 
\eea
\bea
\Lambda_V^{(2)} =    a^2\,\ggcf
&\hspace{-0.15cm}\Biggl(\hspace{-0.15cm}&p^2 \Bigl(0.44724 -0.79717 \,\alpha+c_{\rm sw} (1.83028 \,\omega -1.61663)
-0.14183 \,c_{\rm sw}^2 \nonumber \\
&&\hspace{0.45cm}-1.02045 \,\omega +1.46637 \,\omega^2
+\log(a^2\,p^2) \left(\frac{47}{360}+\frac{3 \,\alpha}{16}+\frac{5\,c_{\rm sw}}{8}-\frac{c_{\rm sw}^2}{8}-\frac{\omega }{2}\right)
     \Bigr) \nonumber \\
&+&
\frac{p4}{p^2} \,\Bigl(2.54053+0.23958 \,\alpha+c_{\rm sw}(0.28454 -0.36251 \,\omega )
+0.12838\,c_{\rm sw}^2 \nonumber \\
&&\hspace{0.45cm}-4.08165 \,\omega -16.0889 \,\omega^2
-\frac{157}{180} \log(a^2\,p^2)\Bigr) \Biggr) \\ [2ex]
\Lambda_A^{(2)} =    a^2\,\ggcf &\hspace{-0.15cm}\Biggl(\hspace{-0.15cm}& p^2 \Bigl(-0.37692-0.07942 \,\alpha+c_{\rm sw}(0.30976 -0.81851 \,\omega )
-0.85384 \,c_{\rm sw}^2 \nonumber \\
&&\hspace{0.45cm}+1.45079 \,\omega +2.40868 \,\omega^2
+\log (a^2\,p^2) \left(\frac{47}{360}-\frac{\alpha}{16}-\frac{c_{\rm sw}}{8}+\frac{5 \,c_{\rm sw}^2}{8}-\frac{\omega}{2}\right)  \Bigr) \nonumber \\
&+&
\frac{p4}{p^2}\, \Bigl(2.54053 +0.23958 \,\alpha + c_{\rm sw} (0.28454 -0.36251 \,\omega )
+0.12838\,c_{\rm sw}^2 \nonumber \\
&&\hspace{0.45cm}-4.08165 \,\omega -16.0889 \,\omega^2
-\frac{157}{180} \log(a^2\,p^2) \Bigr) \Biggr)
\eea
\bea
\Lambda_T^{(2)} =    a^2\,\ggcf &\hspace{-0.15cm}\Biggl(\hspace{-0.15cm}& p^2 \Bigl(0.17468-0.59766 \,\alpha +c_{\rm sw} (1.38881 \,\omega -1.29556)
-0.51102 \,c_{\rm sw}^2 \nonumber \\
&&\hspace{0.45cm}+1.32727 \,\omega  +2.66032 \,\omega^2
+\log(a^2\,p^2) \left(\frac{19}{120}+\frac{\alpha}{8}+\frac{c_{\rm sw}}{2}+\frac{c_{\rm sw}^2}{4}-\omega \right)\Bigr)\nonumber\\
&+&
\frac{p4}{p^2}\, \Bigl(2.85187+0.14583 \,\alpha +c_{\rm sw} (0.28454 -0.36251 \,\omega )
+0.12838\,c_{\rm sw}^2\nonumber \\
&&\hspace{0.45cm}-4.08165 \,\omega-16.0889 \,\omega^2 -\frac{157}{180} \log(a^2\,p^2)\Bigr)\Biggr)
\eea

\section{$\beta-$function and anomalous dimensions}
\label{appB}

In this Appendix we give the definitions of the $\beta-$function
and the anomalous dimension for the fermion field and the local
operators. The perturbative coefficients up to three loops are given
in $SU(3)$ and in the Landau gauge.

The scale dependence of the renormalized operator is encoded
in the anomalous dimension and is defined as:
\be
\gamma^{\mathcal S} = - \mu \frac{\mathrm d}{\mathrm d \mu}
\log Z_{\mathcal S} =
\gamma_0 \frac{g^{\mathcal S} (\mu)^2}{16 \pi^2}
 + \gamma_1^{\mathcal S}
       \left( \frac{g^{\mathcal S} (\mu)^2}{16 \pi^2} \right)^2
 + \gamma_2^{\mathcal S}
       \left( \frac{g^{\mathcal S} (\mu)^2}{16 \pi^2} \right)^3
 + \cdots
\ee
where $\mathcal S$ is the renormalization scheme. The $\beta-$function
is defined as:
\be
\beta^{\mathcal S} =  \mu \frac{\mathrm d}{\mathrm d \mu}
                      g^{\mathcal S} (\mu) =
 - \beta_0 \frac{g^{\mathcal S} (\mu)^3}{16 \pi^2}
 - \beta_1 \frac{g^{\mathcal S} (\mu)^5}{(16 \pi^2)^2}
 - \beta_2^{\mathcal S} \frac{g^{\mathcal S} (\mu)^7}{(16 \pi^2)^3}
 + \cdots\,.
\ee
The coefficients of the $\beta-$function in the ${\overline{\rm MS}}$
  and the RI$'$-MOM schemes coincide up to three loops and are given by
  \cite{vanRitbergen:1997va,Gracey:2003yr}:
\bea
\beta_0 & = & 11 - \frac{2}{3} N_f \,, \\
\beta_1 & = & 102 - \frac{38}{3} N_f \,, \\
\beta_2 & = & \frac{2857}{2} - \frac{5033}{18} N_f
              + \frac{325}{54} N_f^2 \,.
\eea
The coefficients of the anomalous dimension for the quark field in the
${\overline{\rm MS}}$ and RI$'$-MOM schemes are~\cite{Chetyrkin:1999pq}:
\bea
\gamma_0 & = & 0 \,, \\
\gamma_1 & = & \frac{134}{3} - \frac{8}{3} N_f \,, \\
\gamma_2^{\overline{\rm MS}} & = & \frac{20729}{18} - 79 \zeta_3 - \frac{1100}{9} N_f
               + \frac{40}{27} N_f^2 \,,\\
\gamma_2^{{\mbox{\scriptsize RI$^{\prime}$-MOM}}} & = & \frac{52321}{18} - 79 \zeta_3 - \frac{1100}{9}
N_f + \frac{40}{27} N_f^2\,,
\eea
for the scalar/pseudoscalar operators~\cite{Chetyrkin:1997dh,Vermaseren:1997fq}:
\bea
\gamma_0 & = & -8 \,, \\
\gamma_1^{\overline{\rm MS}} & = & - \frac{404}{3} + \frac{40}{9} N_f \,, \\
\gamma_1^{\mbox{\scriptsize RI$^{\prime}$-MOM}} & = & - 252 + \frac{104}{9} N_f \,, \\
\gamma_2^{\overline{\rm MS}} & = & - 2498 + \left( \frac{4432}{27}
           + \frac{320}{3} \zeta_3 \right) N_f + \frac{280}{81} N_f^2 \,,\\
\gamma_2^{\mbox{\scriptsize RI$^{\prime}$-MOM}} & = & - \frac{40348}{3} + \frac{6688}{3}\zeta_3 +\left( \frac{35176}{27}
           - \frac{256}{9} \zeta_3 \right) N_f - \frac{1712}{81} N_f^2 \,,
\eea
($\zeta_3 = 1.20206...$) and for the tensor~\cite{Gracey:2000am,Gracey:2003yr}:
\bea
\gamma_0 & = & \frac{8}{3} \,, \\
\gamma_1 & = & \frac{724}{9} - \frac{104}{27} N_f \,, \\
\gamma_2^{\overline{\rm MS}} & = & \frac{105110}{81} - \frac{1856}{27} \zeta_3
             - \left( \frac{10480}{81}
             + \frac{320}{9} \zeta_3 \right) N_f - \frac{8}{9} N_f^2\,,\\
\gamma_2^{\mbox{\scriptsize RI$^{\prime}$-MOM}} & = & \frac{359012}{81} - \frac{26144}{27} \zeta_3
             + \left(-\frac{39640}{81}
             + \frac{512}{27} \zeta_3 \right) N_f + \frac{2288}{243} N_f^2\,.
\eea

\section{An alternative for the pion pole subtraction}
\label{appC}
Here we mention an alternative in handling the pion pole subtraction to determine 
the renormalization factors for $Z_P$. 
We can perform a global fit to the ratio of Eq. (\ref{fit_ratio})
taking into account the data at all pion masses and all scales
simultaneously. Since in the same fit we combine both correlated and
uncorrelated data extracted from different ensembles, we use the
super-jackknife procedure~\cite{Bratt:2010jn} for the error estimation
of the fit parameters. We employ a fit function with four, five and six
parameters, of the form:
\bea
\label{f4}
f^{(4)}(p,\,m_\pi) &=& a_0 + a_2\,p^2 + \frac{c_0 +
  c_2\,p^2}{m_\pi^2}\,,\\[1ex]
\label{f5}
f^{(5)}(p,\,m_\pi) &=& a_0 + a_2\,p^2 + b_0\,m_\pi^2 +
\frac{c_0 + c_2\,p^2}{m_\pi^2}\,,\\[1ex]
\label{f6}
f^{(6)}(p,\,m_\pi) &=& a_0 + a_2\,p^2 + (b_0 + b_2\,p^2)\,m_\pi^2 +
\frac{c_0 + c_2\,p^2}{m_\pi^2}\,.
\eea
In the above functions the parameters $a_i,\,b_i,\,c_i$ are
constants, but their estimation depends on the range of momenta that
we take into account. Since the data at $(a\,p)^2<1$ do not exhibit
any plateau behavior, they are entirely excluded from the fit. In fact, in the
application of each fit shown in Eqs.~(\ref{f4}) - (\ref{f6}) we use
data corresponding to various momentum ranges, $(a\,p)^2$ (see text below). 
Similarly to the case of the two- and three-parameter fit, we are interested in extracting $Z_P$ in
${\overline{\rm MS}}$ with a single fit which is related to $a_0$,
through $Z_P^{\overline{\rm MS}}=(a_0)^{-1}$. To summarize, we have a
total of 12 estimates for $Z_P^{\overline{\rm MS}}$ extracted from
the global fits, which correspond to all combinations between the three
fit functions (Eqs.~(\ref{f4}) - (\ref{f6})) and the four momentum ranges in $(a\,p)^2 \in [1:10]$, $[2:6]$, $[2:10]$ and $[3.7-10]$.
We find that the six-parameter fit is not very stable, while
the four- and five-parameter global fits give compatible results with the 
local two- and
three-parameter local fits applied to each momentum squared
individually. An interesting observation is that the results obtained
using the interval $[2:10]$ are compatible for all types of fits
discussed here (see Table~\ref{tab_ZP_4_6}). 
The same global fitting has been tested to find $Z_P/Z_S$.
\begin{table}[!h]
\begin{center}
\begin{tabular}{|c|c|c|c|c|}
\hline
$\,\,$ &4-parameter fit & 5-parameter fit  & 6-parameter fit \\
\hline
$Z_P^{\overline{\rm MS}}$ & 0.4917(022)    &0.4921(078)    &0.4852(176) \\
\hline
\end{tabular}
\end{center}
\caption{Estimates for $Z_P$ in the ${\overline{\rm MS}}$ scheme using
  the global fits of Eq.~(\ref{f4}) - (\ref{f6}) and the data in the
  momentum range $[2:10]$. Statistical errors are shown in parentheses.}
\label{tab_ZP_4_6}
\end{table}
\FloatBarrier

We find that the four-, five- and six-parameter global fits are, in general, more
unstable compared to the two- and three-parameter local fits and will not be used
as final estimates. Moreover, the restriction of the global fitting
functions to have quadratic dependence with respect 
to the momentum (Eqs.~(\ref{f4}) - (\ref{f6})) is not based on any
theoretical arguments. Nevertheless, agreement within errorbars
between different fits gives confidence on the final estimate.

As r\'{e}sum\'{e} of these studies we use in Sec.~\ref{sec5} 
the local two-parameter fit to remove the pion pole and combine it with a
linear fit in the $(a\,p)^2$ interval $[2:10]$ counting for a remaining
momentum dependence in the non-pole term after chiral
extrapolation. The resulting constant is then the inverse of the
renormalization function in the ${\overline{\rm MS}}$ scheme.

\bibliographystyle{apsrev}  

\bibliography{Zfactors_local_bib}

\end{document}